\documentclass[final,5p,times]{elsarticle}
\usepackage{graphicx}
\usepackage{epsfig}
\usepackage{dcolumn}
\usepackage{bm}
\usepackage{amsmath}
\usepackage{amsfonts}
\usepackage[usenames,dvipsnames]{color}
\usepackage{ulem}

\allowdisplaybreaks

\newcounter{bla}

\journal{Computer Physics Communications}

\renewcommand{\vec}[1]{{\mathbf{#1}}}

\newcommand{\nn}{\nonumber}

\newcommand{\vnabla}{\boldsymbol{\mathbf\nabla}}
\newcommand{\vsigma}{\boldsymbol{\mathbf\sigma}}

\newcommand{\sigmavec}{\boldsymbol{\mathbf\sigma}}

\newcommand{\be}{\begin{equation}}
\newcommand{\ee}{\end{equation}}
\newcommand{\bea}{\begin{eqnarray}}
\newcommand{\eea}{\end{eqnarray}}
\newcommand{\ba}{\begin{array}}
\newcommand{\ea}{\end{array}}
\newcommand{\bc}{\begin{center}}
\newcommand{\ec}{\end{center}}

\definecolor{orange}{RGB}{255,127,0}

\definecolor{celestialblue}{rgb}{0.29, 0.59, 0.82}
\definecolor{cerisepink}{rgb}{0.93, 0.23, 0.51}
\definecolor{coquelicot}{rgb}{1.0, 0.22, 0.0}
\definecolor{darkorchid}{rgb}{0.6, 0.2, 0.8}
\definecolor{emerald}{rgb}{0.31, 0.78, 0.47}
\definecolor{mbscolor}{rgb}{0.31, 0.78, 0.47}

\setcitestyle{numbers}

\begin{document}

\begin{frontmatter}

\title{Solution of the Skyrme HF+BCS equation on a 3D mesh\\
       II. A new version of the \texttt{Ev8} code}

\author[ulb]{W. Ryssens}

\author[ulb]{V. Hellemans}

\author[cenbg1,cenbg2]{M. Bender}

\author[ulb]{P.-H. Heenen\corref{cor1}}
\ead{phheenen@ulb.ac.be}\cortext[cor1]{Corresponding author}

\address[ulb]{PNTPM, CP229,
             Universit{\'e} Libre de Bruxelles,
             B-1050 Bruxelles,
             Belgium}
\address[cenbg1]{
             Universit\'e de Bordeaux, Centre d'Etudes Nucl\'eaires de
             Bordeaux Gradignan, UMR5797, F-33175 Gradignan, France}
\address[cenbg2]{
             CNRS/IN2P3, Centre d'Etudes Nucl{\'e}aires de
             Bordeaux Gradignan, UMR5797, F-33175 Gradignan, France}

\date{8 May 2014}

\begin{abstract}
We describe a new version of the \texttt{Ev8} code that solves the nuclear Skyrme-Hartree-Fock+BCS problem using a 3-dimensional cartesian mesh.
Several new features have been implemented with respect to the earlier version published in 2005. In particular, the numerical accuracy has been improved for a given mesh size by
(i) implementing a new solver to determine the Coulomb potential for protons (ii) implementing a more precise method to calculate the derivatives on a mesh that had already been implemented earlier in our beyond-mean-field codes. The code has been made very flexible to enable the use of a large variety of Skyrme energy density functionals that have been introduced in the last years. Finally, the treatment of the constraints that can be introduced in the mean-field equations has been improved. The code
\texttt{Ev8} is today the tool of choice to study the variation of the energy of a nucleus from its ground state to very elongated or triaxial deformations with a well-controlled accuracy.

\end{abstract}


\end{frontmatter}


{\bf PROGRAM SUMMARY/NEW VERSION PROGRAM SUMMARY}

\begin{small}
\noindent
{\em Manuscript Title:} \\
Solution of the Skyrme HF+BCS equation on a 3D mesh      \\
  II. A new version of the \texttt{Ev8} code             \\
{\em Authors:} W. Ryssens, V. Hellemans, M. Bender, P.-H. Heenen  \\
{\em Program Title: \texttt{Ev8}}                             \\
{\em Journal Reference:}                                      \\
{\em Catalogue identifier:}                                   \\
{\em Licensing provisions:} None                              \\
{\em Programming language: FORTRAN-90}                        \\
{\em Computer:} AMD Opteron 6274, AMD Opteron 6134, AMD Opteron 2378, Intel Core i7-4700HQ\\                                            \\
{\em Operating system:}  Unix, Linux, OS X\\
{\em RAM:} On the order of 64 megabytes for the examples provided.  
{\em Classification:}  17.22: Hartree-Fock Calculations        \\
{\em Catalogue identifier of previous version:}  ADWA          \\
{\em Journal reference of previous version:}                   \\
 P. Bonche, H. Flocard, P.-H. Heenen, CPC \textbf{171} (2005), 49-62\\
{\em Does the new version supersede the previous version?:} Yes, but when used in the same conditions both codes give the same results. \\
{\em Nature of problem:}\\
 By means of the Hartree--Fock + BCS method for Skyrme-type energy density functionals, \texttt{Ev8} allows to
study of the evolution of the binding energy of even-even atomic nuclei for various shapes determined by the most general quadrupole and monopole constraints.
   \\
{\em Solution method:}\\
The program expands the single-particle wave-functions on a 3D Cartesian mesh. The nonlinear mean-field
equations are solved by the imaginary time step method. A quadratic constraint is used to obtain states corresponding to given values of the monopole and quadrupole operators.
   \\
{\em Summary of revisions:}
 \begin{enumerate}
  \item Skyrme energy functionals with tensor terms
  \item Improved accuracy for calculating derivatives
  \item Improved accuracy for solving Coulomb problem
  \item Improvement of the numerics of constraints
 \end{enumerate}
{\em Restrictions:}\\
  {\tt Ev8} assumes time-reversal invariance and nuclear shapes exhibiting three plane-reflection symmetries. Pairing correlations are treated at the BCS level of approximation. \\
{\em Running time:} A few minutes for the examples provided, which concern rather heavy nuclei in modest boxes with an initial guess of Nilsson wavefunctions.\\

\end{small}

%
%
\section{Introduction}
At present, the only microscopic theoretical tools that can be applied throughout the entire nuclear
chart are methods based on a nuclear energy density functional (EDF). Among these, the self-consistent
mean-field approach is the simplest one and can be used as a starting point to introduce correlations beyond
the mean field. Three main families of EDFs are extensively used for low-energy nuclear spectroscopy~\cite{bender03a}.
One of them, the Skyrme EDF, is built from local densities and is tailor-fit for numerical schemes that solve the
mean-field equations in coordinate space.

Although frequently called Skyrme--Hartree--Fock (Skyrme-HF) approach,
the method described here is also often labeled as
density-dependent HF,
HF+BCS,
self-consistent mean-field,
nuclear density functional theory (DFT), or
single-reference energy density functional (SR EDF) method
in the literature.
Each of these designations tries to underline that the effective interaction
is used in a manner that in one way or the other differs from what is
described as HF equations in introductory textbooks, although the
equations that are solved are similar.
First of all, the density dependence of Skyrme's interaction gives
rise to rearrangement terms \cite{blaizotripka,RingSchuck}
that are absent in ordinary HF.
Second, in the code described here, pairing correlations are taken into
account in the BCS approximation to the full
Hartree--Fock--Bogoliubov scheme which simplifies the equations to be
solved.
Third, and most importantly, in this method the total energy is rarely
set-up as the expectation value of a many-body Hamiltonian. On the one
hand, the pairing interaction is chosen to be different from the
Skyrme and Coulomb interactions used for the ``particle-hole" or
``mean-field" part. On the other hand, specific terms in both the Skyrme
and Coulomb energies are also routinely modified.
We will come back to this in Sect.~\ref{sect:EDF:skyrme:func}.

The motivations for these distinctions and their often subtle formal
consequences are irrelevant for the purpose of the present paper and
we refer to Refs.~\cite{bender03a,erler11a,sadoudi13a,lac09a} for details.
The only relevant point for our discussion is that the fundamental
object in the present framework is the expression for the total
energy, the EDF, and not a many-body Hamiltonian.

In this article, we present an update of the \texttt{Ev8} code first published in Ref.~\cite{BFH05a}
to solve the mean-field equations for the Skyrme EDF. Since 1985 \cite{BFH85a}, a large number
of problems in nuclear spectroscopy -- going from the lightest to the
heaviest nuclei -- has been addressed. This code has also
been the starting point for beyond-mean-field calculations based on angular-momentum projection
and configuration mixing in the Generator Coordinate Method~\cite{BH08a,BBH08a}.

In \texttt{Ev8}, the single-particle wave functions are discretized on a 3-dimensional (3D) mesh to solve the mean-field equations.
This gridding technique was first introduced in the 1970's to solve the time-dependent Hartree-Fock (HF) problem
\cite{cusson76a,flocard78a,bonche78a}, for which it is still the representation of choice today~\cite{maruhn,simenel12a}.
Such a coordinate-space representation offers several advantages over the widely-used harmonic-oscillator (HO) basis
representation \cite{HFODD1,HFODD8,HFBTHO1,HOSPHE1}. Indeed, the single-particle states exhibit
the correct asymptotic behavior by construction whereas a rescaling  \cite{HFBTHO1} is required for an HO expansion.
Moreover, a large variety of different shapes associated with nuclear states can be described with similar accuracy.
This has to be contrasted with the need to optimize the parameters of the HO basis for each deformation  or to introduce a two-centre basis for
very elongated shapes. It comes at the prize that coordinate-space representations are more demanding
than a HO expansion from a
computational point of view.

Since the 1970s, several implementations of the equidistant discretization of space have been proposed in the literature and
they mainly differ by the symmetries imposed on the mean-field wave function and the treatment of derivative operators.
For the latter, the most straightforward choice are the finite-difference formulas which are adopted in several 1D spherical codes
\cite{reinhard91a,HFBRAD1,Karim} and in 3D codes \cite{BFH05a,BFH85a}. Alternative choices for the 3D codes are
a Fourier representation of the derivatives \cite{maruhn,blum92a} or splines-based techniques \cite{blum92a}. Finally,
one can opt for a Lagrange mesh representation as proposed in~\cite{BH86a}. This representation is a subclass of the so-called
\textit{Discrete Variable Representation} in Quantum Chemistry \cite{DVR,bulgac13a}. In short, the spatial functions are expanded
 onto an orthogonal set of continuous basis functions, each of which is non-zero at only one of the collocation points.
In the Lagrange-mesh method, the basis functions are chosen such that the derivatives at the mesh points are the exact inverse
of the integration. The same technique applied to the rotation
operators for angular-momentum projection \cite{BH08a,VHB00a} leads to
very accurate results, independent of the size of the rotation angles.

In \texttt{Ev8}, we impose time-reversal invariance to the wave function and spatial
symmetries are chosen such that the space can be limited to one octant of the box. This
significantly reduces the computing time and enables one to perform large-scale
calculations, even with rather limited computer resources. In addition to \texttt{Ev8},
our group has set up a family of codes where various combinations of symmetries are
lifted and that will also be published in the future.
By construction, \texttt{Ev8} can be used to study triaxial deformation in the ground state
of nuclei. It also permits to verify whether minima obtained
in axial calculations are stable against $\gamma$-de\-for\-ma\-tions.

Compared to the previous version of the code, the most important updates concern

\begin{itemize}



\item
The treatment of the various options and choices for the EDF. All Skyrme parametrizations
widely used today are based on the same form
of the EDF. During their fit, however, different choices have been made for the treatment of
the density-dependent, spin-orbit, and tensor terms of the Skyrme interaction,
the Coulomb exchange term,
the corrections for spurious motion of the centre of mass,
and also for the values of fundamental constants such as the nucleon
masses.
The updated \texttt{Ev8} offers most of these options, as parametrizations
should always be used with the original choices made for their fit.

\item The tensor terms of the EDF have been introduced in the code and their
different representations occurring in the literature are printed.

\item
A new numerical representation of the derivatives of functions on a 3D mesh has been implemented~\cite{BH86a}. This treatment was already included in our beyond-mean-field codes~\cite{BH08a,VHB00a}
and greatly improves the numerical accuracy of \texttt{Ev8}.

\item
The solution of the Coulomb potential on the 3D mesh is now based on a second-order discretisation of the Laplacian operator,
resulting in a significant gain of accuracy.

\item
New cutoff procedures for the constraints - mainly the quadrupole constraint - as well as a new constraint have been incorporated. In practice, the cutoff proposed by Rutz \textit{et al.} ~\cite{rutz95a} is found to be the most stable.

\item
The in- and output of the code were reorganized and extended to make them more transparent.

\end{itemize}

In the following, we limit ourselves to the presentation of the key equations
necessary to understand the input and the output of the code, most of which
have already been presented in other publications
to which we refer the reader for further details.


\section{Principles of the method}

Here, we briefly review those aspects of the self-consistent mean-field
approach relevant to the discussion of the features of \texttt{Ev8}.

\subsection{The many-body state}

The many-body wave function of a nucleus is determined using the HF+BCS approximation.
In short, one assumes that the single-particle Hamiltonian $\hat{h}$ and the
density matrix $\hat{\rho}$ can be diagonalized simultaneously, or, equivalently,
that only the pairing matrix elements between pairs of conjugate states are
different from zero~\cite{RingSchuck}.
The paired independent-particle state then takes the form
\begin{equation}
\label{eq:BCS}
| \text{BCS} \rangle
= \prod_{k > 0} ( u_k + v_k a^\dagger_k a^\dagger_{\bar{k}} ) \, | 0 \rangle
  \, ,
\end{equation}
where $v_k^2$ are the occupation probabilities of the single-particle
states $\Psi_k(\vec{r})$ and $\Psi_{\bar{k}}(\vec{r})$.
The $u_k$ and $v_k$ are linked by the normalization condition
\mbox{$u^2_k + v^2_k = 1$}. We use the phase convention
\mbox{$u_k = u_{\bar{k}} \geq 0$} and \mbox{$v_k = -v_{\bar{k}} \geq 0$}.

\subsection{Single-particle states}
\label{sect:symmetries}

In a Cartesian 3D representation, the single-particle states $\Psi_k(\vec{r}) = \langle \vec{r} | a^\dagger_k | 0 \rangle$ are
represented by four real functions corresponding to the real and
imaginary parts of the upper and lower spinor components
\begin{equation}
\Psi_k (\vec{r})
  =   \left( \begin{array}{c}
             \psi_{k} (\vec{r},\sigma = +) \\
             \psi_{k} (\vec{r},\sigma = -)
             \end{array}
      \right)
  =   \left( \begin{array}{c}
             \psi_{k,1} (\vec{r}) + \textrm{i} \, \psi_{k,2} (\vec{r}) \\
             \psi_{k,3} (\vec{r}) + \textrm{i} \, \psi_{k,4} (\vec{r})
             \end{array}
      \right) \, .
\end{equation}
The \texttt{Ev8} code is restricted to nuclear shapes that have three plane reflection
symmetries in the \mbox{$x = 0$}, \mbox{$y = 0$}, and \mbox{$z = 0$}
planes. This allows to reduce the calculation to $1/8$ of
the full box
\begin{figure}[t!]
\label{Boxes}
\centerline{\includegraphics[width=6cm]{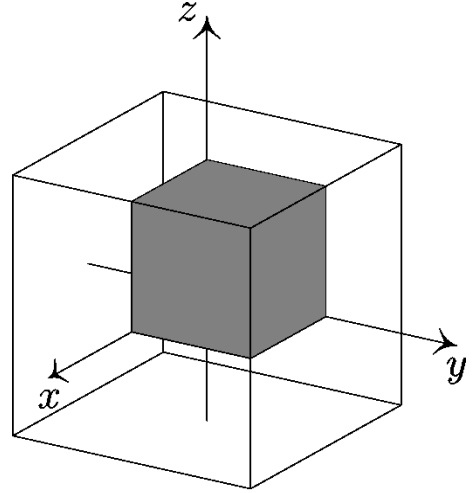}}
\caption{The symmetries assumed allow \texttt{Ev8} to represent the nucleus only in 1/8th
of the total box, indicated by the grayed cube.}
\end{figure}
and can be achieved by choosing the single-particle wave functions
to be irreps of the three-generator subgroup of the $D^{DT}_{2h}$
group~\cite{dob00a}, generated by
$z$ signature $\hat{R}_z$, parity $\hat{P}$ and the $y$ time simplex
$\hat{S}_y^{T}$
\begin{subequations}
\begin{eqnarray}
\hat{R}_z \, \Psi_k (\vec{r})
& = & \text{i} \, \eta_k \, \Psi_k (\vec{r}) \, ,
      \\
\hat{P} \, \Psi_k (\vec{r})
& = & \pm \Psi_k (\vec{r}) \, ,
      \\
\hat{S}_y^{T} \, \Psi_k (\vec{r})
& = & + \Psi_k (\vec{r}) \, .
\end{eqnarray}
\end{subequations}
The wave functions $\Psi_k(\vec{r})$ themselves, however, do not acquire
plane reflection symmetries. Instead, the real functions
$\psi_{k,j} (\vec{r})$ have a specific plane reflection symmetry
depending on the signature and
parity of the single-particle state, cf.\ Ref.~\cite{hellemans12a} for details.
The anti-linear $\hat{S}_y^{T}$ operator not only establishes a spatial
symmetry of the single-particle states, but also fixes their relative
phases \cite{dob00a,dob00b}.

In addition to these spatial symmetries, the single-particle states are imposed to be
pure proton or neutron states. Time-reversal
invariance is also enforced, by requiring that
single-particle states of opposite signature are related by time-reversal,
$\Psi_{\bar{k}}(\vec{r}) = \hat{T} \, \Psi_k(\vec{r})$ and
$\Psi_k(\vec{r}) = -\hat{T} \, \Psi_{\bar{k}}(\vec{r})$, and form two-fold
Kramers-degenerate pairs. Such pairs of single-particle states will be called
\textit{conjugate states}, where we use the usual convention that
\mbox{$k > 0$} labels single-particle states $\Psi_k(\vec{r})$ of positive
signature $\eta_k$ and {\mbox{$\bar{k} < 0$} single-particle states
of negative signature $\eta_{\bar{k}}$. Only positive signature states and
their properties are explicitly calculated and printed by the code.

\subsection{The local densities}
\label{sect:densities}

With the symmetries assumed here, only a few local normal densities
\begin{eqnarray}
\rho_q (\vec{r})
& = & 2 \sum_{k > 0} v^2_k \, \Psi^\dagger_k (\vec{r}) \,
                     \Psi_k (\vec{r}) \, ,
      \\
\tau_q (\vec{r})
& = & 2 \sum_{k > 0} v^2_k \, \big[ \vnabla \Psi_k (\vec{r}) \big]^\dagger
                     \cdot \big[ \vnabla \Psi_k (\vec{r}) \big]
      \\
J_{q,\mu \nu}(\vec{r})
& = & - \textrm{i} \sum_{k > 0} v^2_k \,
      \Big\{  \Psi^\dagger_k (\vec{r}) \, \hat{\sigma}_\nu \,
              \big[ \nabla_\mu \Psi_k (\vec{r}) \big]
      \nn \\
&   & \qquad \quad
             -\big[ \nabla_\mu \Psi_k (\vec{r}) \big]^\dagger \,
              \hat{\sigma}_\nu \, \Psi_k (\vec{r})
      \Big\}
      \, ,
\end{eqnarray}
for protons and neutrons, \mbox{$q = p$}, $n$, are needed to calculate the
total energy and other observables. These are the local density $\rho$,
the kinetic density $\tau$, and the Cartesian components of the
spin-current tensor density $J_{\mu \nu}$. For the expressions of the
densities in terms of the real functions $\psi_{k,i}$ we refer to
the Appendices of Ref.~\cite{hellemans12a}, which also discuss their plane
reflection symmetries. A detailed
discussion of the relation between these local densities and the full
one-body density matrix and of their general properties is provided by
 Ref.~\cite{dob00a}.

Finally, instead of the Cartesian components of the spin-current density,
it is sometimes preferable to recouple them to a pseudo-scalar
$J^{(0)}_q (\vec{r})$, an antisymmetric vector $J^{(1)}_{q, \kappa} (\vec{r})$
and a symmetric and traceless pseudo-tensor $J^{(2)}_{q, \mu \nu} (\vec{r})$
\cite{per04a,ben09a}
\begin{equation}
\label{eq:J:spherical}
J_{q,\mu \nu}(\vec{r})
 = \tfrac{1}{3} \, \delta_{\mu \nu} \, J^{(0)}_q(\vec{r})
   + \tfrac{1}{2} \sum_{\kappa} \epsilon_{\mu \nu \kappa} \,
     J^{(1)}_{q, \kappa} (\vec{r})
   + J^{(2)}_{q, \mu \nu} (\vec{r}) \, ,
\end{equation}
where $\delta_{\mu \nu}$ is the Kronecker symbol and
$\epsilon_{\mu \nu \kappa}$ the Levi-Civita tensor. The inverse
relations are given by
\begin{eqnarray}
\label{eq:J0}
J^{(0)}_q (\vec{r})
& = & \sum_{\mu} J_{q, \mu \mu}(\vec{r}) \, ,
      \\
\label{eq:J1}
J^{(1)}_{q, \kappa} (\vec{r})
& = & \sum_{\mu \nu} \epsilon_{\kappa \mu \nu} \, J_{q, \mu \nu}(\vec{r})
      \, , \\
\label{eq:J2}
J^{(2)}_{q, \mu \nu} (\vec{r})
& = & \tfrac{1}{2} \big[ J_{q, \mu \nu}(\vec{r}) + J_{q, \nu \mu}(\vec{r}) \big]
      - \tfrac{1}{3} \delta_{\mu \nu} \! \!
        \sum_{\kappa} J_{q, \kappa \kappa}(\vec{r}) \, .
\end{eqnarray}
For conserved parity as assumed here, the pseudo-scalar density
$J^{(0)}_q(\vec{r})$ is zero.

With the symmetries chosen here, a proton-neutron representation is the
most efficient from a numerical implementation point of view. For the discussion of
the physics content of the EDF on the other hand, it is more advantageous to
recouple the proton and neutron densities to isoscalar ($t=0$) and
isovector ($t=1$) densities, i.e.
\begin{eqnarray}
\label{eq:rho:iso}
\rho_0 (\vec{r})
& = & \rho_n (\vec{r}) + \rho_p (\vec{r}) \, ,
      \nn \\
\rho_1 (\vec{r})
& = & \rho_n (\vec{r}) - \rho_p (\vec{r}) \, ,
\end{eqnarray}
and similar for $\tau(\vec{r})$ and $J_{\mu \nu}(\vec{r})$.

%
%

\subsection{Mean-field and pairing equations}
\label{sect:HF+BCS}

The total binding energy is
given by the sum of the kinetic energy, the Skyrme EDF
that models the effective interaction between nucleons, the Coulomb
energy, the pairing energy, and corrections for spurious motions
\begin{equation}
E
=   E_{\rm kin}
  + E_{\rm Sk}
  + E_{\rm Coul}
  + E_{\rm pair}
  + E_{\rm corr}
  \label{eq:EF:schematic}
\end{equation}
Different interactions are used for the mean-field and the pairing channels.
Hence, the contributions of both channels to the total energy are separated,
which makes our approach not a HF+BCS approximation \textit{in stricto sensu}.

In practice, this means that the energy is separated into a one-body
kinetic contribution, a particle-hole type many-body part consisting of $E_{\rm Sk}$
and $E_{\rm Coul}$, and a pairing part
\begin{equation}
E
= E_{\text{kin}} + E_{\text{ph}} + E_{\text{pair}} \, ,
\end{equation}
which in the canonical basis can be written as
\begin{eqnarray}
E_{\text{kin}}
& = & 2 \sum_{k > 0} v_k^2 \, t_{kk} \, ,
      \\
E_{\text{ph}}
& = & 4 \sum_{k, m > 0} v_k^2 \, v_m^2 \, v^{\text{ph}}_{km km} \, ,
      \\
\label{eq:epair}
E_{\text{pair}}
& = & \sum_{k, m > 0} f_k \, u_k v_k \, f_m \, u_m v_m \,
      \bar{v}^{\text{pair}}_{k \bar{k} m \bar{m}}
\, .
\end{eqnarray}
The $t_{kk}$ are matrix elements of the kinetic energy, the
$v^{\text{ph}}_{km km}$ are (non-antisymmetrized) matrix elements
characterizing the mean-field or particle-hole interaction,
whereas the $\bar{v}^{\text{pair}}_{k \bar{k} m \bar{m}}$ are anti-symmetrized matrix
elements of a pairing interaction and the $f_i$ are cutoff factors that
will be specified later.
Depending on their nature, some correction terms $E_{\rm corr}$ will
be treated as being part of $E_{\text{ph}}$, others as being part of
$E_{\text{pair}}$.

As it will be discussed later, in practice,
the matrix elements $v^{\text{ph}}_{km km}$ are most of the time not related to an interaction.

The equations of motion are derived from the condition of a stationary
total energy under the restriction of orthogonal single-particle states, and with
additional constraints on the particle numbers $N$ and $Z$
\cite{blaizotripka,RingSchuck}.
This leads to two systems of coupled equations. Those for the single-particle states are given by
\begin{eqnarray}
\label{eq:hf:equation}
\hat{h}_q (\vec{r}) \, \Psi_k (\vec{r})
& = & \epsilon_k \, \Psi_k (\vec{r}) \, .
\end{eqnarray}
With the symmetries assumed here, the kinetic, Skyrme and Coulomb energies
can be rewritten as integrals over energy densities that are a
functional of the local densities
$\rho_q(\vec{r})$, $\tau_q(\vec{r})$, and $J_{q,\mu \nu}(\vec{r})$
\begin{equation}
E_{\rm i}
= \int \! d^3 r \;
  \mathcal{E}_{\rm i} \big[ \{ \rho_q, \tau_q, J_{q,\mu \nu} \} \big] \, .
\end{equation}
The single-particle Hamiltonian is then found to be \cite{hellemans12a}
\begin{equation}
\label{eq:hf:h}
\hat{h}_q (\vec{r})
 = - \vnabla \cdot B_q (\vec{r}) \, \vnabla
      + U_q (\vec{r})
      - \textrm{i} \sum_{\mu \nu}
       W_{q,\mu \nu} (\vec{r}) \, \nabla_\mu \, \hat{\sigma}_\nu
\, ,
\end{equation}
where $B_q (\vec{r}) = \hbar^2 / 2 m_q^*(\vec{r}) = \delta E / \delta \tau_q(\vec{r})$
is proportional the inverse of a po\-si\-tion-dependent effective mass,
$U_q (\vec{r}) = \delta E / \delta \rho_q(\vec{r})$ is the central potential,
and $W_{q,\mu \nu} (\vec{r}) = \delta E / \delta J_{q,\mu \nu} (\vec{r})$
a generalization of the spin-orbit potential in the presence of
tensor terms.

The set of equations that determine the occupation numbers $v^2_k$ of
the single-particle states of each nucleon species $q$ are derived from
the variation of
\begin{equation}
\label{eq:BCS:2}
\frac{\delta}{\delta v_j}
\Big( 2 \sum_{k > 0} \epsilon_k v^2_k - \lambda_q \langle \hat{N}_q \rangle
+ E_{\text{pair}} \Big)
= 0 \, ,
\end{equation}
where the $\lambda_q$ are Lagrange multipliers (or the Fermi energies),
introduced to obtain the requested mean number of protons and neutrons
and \mbox{$u_k = \big( 1 - v_k^2 \big)^{1/2}$} is a
function of $v_k^2$.
For a given set of single-particle states and single-particle energies
$\epsilon_k$, determined at each mean-field iteration, this leads to
the expression
\begin{equation}
\label{eq:BCS:v2}
v_k^2
= \frac{1}{2}
  \left[ 1
        - \frac{\epsilon_k - \lambda_q}
               {E_{\text{qp},k}}
  \right]\, ,
\end{equation}
for the occupation numbers for proton and neutron states, where
\begin{equation}
\label{eq:Equasi:BCS}
E_{\text{qp},k}
= \sqrt{ (\epsilon_k - \lambda_q)^2 + f_k^2 \, \Delta^2_{k \bar{k}}} \, ,
\end{equation}
are the quasi-particle energies and $\Delta_{k \bar{k}}$
is the pairing gap for each pair of conjugate single-particle states
\begin{equation}
\label{eq:Delta}
\Delta_{k \bar{k}}
= \sum_{m > 0} f_m \, u_m v_m \, \bar{v}^{\text{pair}}_{k \bar{k} m \bar{m}}
\, .
\end{equation}
The two systems of equations \eqref{eq:hf:equation} and \eqref{eq:BCS:v2}
are coupled by self-consistency and need to be solved for each particle
species.

For further details about the BCS scheme and the interpretation
of the quantities defined above we refer to \cite{RingSchuck,Nilsson,Rowe70a}.


\section{The energy density functional}
\label{sect:EDF}

%
%
\subsection{The kinetic energy}
\label{sect:EDF:kin}

The kinetic energy density is given by
\begin{equation}\label{E_kin}
\mathcal{E}_{\rm kin}
= \sum_{q = n,p} \frac{\hbar^2}{2m_q} \, \tau_q (\vec{r})\, ,
\end{equation}
where the summation runs over protons and neutrons. In the new version
of the code it is possible to set the values for $\hbar^2/(2 m_q)$
to any value, which also might be different for protons and neutrons
(see the explanation of the data in Sect.~\ref{sect:data}).
Only when the two masses are equal, the kinetic energy is
an isoscalar.

%
%
\subsection{The Skyrme energy}
\label{sect:EDF:skyrme}

As explained in the Introduction, there are two main ways to introduce the ``standard''
Skyrme EDF. It can be derived directly from a Skyrme force or be constructed from local
densities and their derivatives in the spirit of DFT. The early parametrizations, proposed
in particular by the Orsay group and still widely used today~\cite{Bei75a,Kri80a,Bar82a},
were determined for a (density-de\-pen\-dent) Skyrme force, but some contributions to
the energy generated by the force were neglected for numerical reasons. Following
the modern terminology they would have to be called EDFs, as most of the more recent ones.
These two different points of
views to introduce EDFs have led to different ways of expressing the parametrizations. In
the following sections, we present how they are related. We discuss in particular the terms
related to the tensor interaction that were not included in the previous version of the code.

\subsubsection{The standard Skyrme force}
\label{sect:EDF:skyrme:v}

Let us start with a Skyrme force that consists of central,
spin-orbit and tensor interactions
\begin{equation}
\hat{v}
=   \hat{v}^{\text{central}}
  + \hat{v}^{\text{tensor}}
  + \hat{v}^{\text{LS}}
\, .
\end{equation}
The most widely-used form of the density-dependent central two-body
Skyrme interaction is defined as \cite{vautherin72a}
\begin{eqnarray}
\label{eq:Skyrme:central}
\lefteqn{\hat{v}^{\text{central}} (\vec{r},\vec{r}')
} \nn \\
& = & t_0 \, ( 1 + x_0 \hat{P}_\sigma ) \; \delta (\vec{r}-\vec{r}')
      \nn \\
&   & + \tfrac{1}{2} \, t_1 \, ( 1 + x_1 \hat{P}_\sigma ) \,
        \Big[   \hat{\vec{k}}^{\prime 2} \; \delta (\vec{r}-\vec{r}')
              + \delta (\vec{r}-\vec{r}') \; \hat{\vec{k}}^2
        \Big]
      \nn \\
&   & + t_2 \ ( 1 + x_2 \hat{P}_\sigma ) \,
      \hat{\vec{k}}^{\prime} \cdot \delta (\vec{r}-\vec{r}') \; \hat{\vec{k}}
      \nn \\
&   & + \tfrac{1}{6} \, t_{3a} \, ( 1 + x_{3a} \hat{P}_\sigma ) \;
        \rho_0^{{\alpha_a}} (\vec{R}) \; \delta (\vec{r}-\vec{r}')
      \nn \\
&   & + \tfrac{1}{6} \, t_{3b} \, ( 1 + x_{3b} \hat{P}_\sigma ) \;
        \rho_0^{\alpha_b} (\vec{R}) \; \delta (\vec{r}-\vec{r}') \, ,
\end{eqnarray}
where $\hat{P}_\sigma$ is the spin exchange operator,
$\hat{\vec{k}} \equiv -\frac{\mathrm{i}}{2}(\vnabla -\vnabla')$ the
relative momentum operator acting to the right
and $\hat{\vec{k}}'$ is its complex conjugate
acting to the left, and $\rho_0 (\boldsymbol{R})$ is the isoscalar density
at $\vec{R} \equiv \tfrac{1}{2} ( \vec{r} + \vec{r}' )$.
The spin-orbit part is given by \cite{vautherin72a,bell56a}
\begin{eqnarray}
\label{eq:Skyrme:LS}
\hat{v}^{\text{LS}} (\vec{r},\vec{r}')
& = & i W_0 \, ( \hat{\sigmavec}_1 + \hat{\sigmavec}_2 ) \cdot
        \hat{\vec{k}}^{\prime} \times
        \delta (\vec{r}-\vec{r}') \;  \hat{\vec{k}}
\end{eqnarray}
which sometimes is also accompanied by a tensor part \cite{sky56a}
\begin{align}
\label{eq:Skyrme:tensor}
\hat{v}^{\text{tensor}} &(\vec{r},\vec{r}')
 \nn \\
 = & \tfrac{1}{2} \; t_e \;
    \Big\{
    \big[ 3 \,( \vsigma_1 \cdot \vec{k}' ) \, ( \vsigma_2 \cdot \vec{k}' )
          - ( \vsigma_1 \cdot \vsigma_2 ) \, \vec{k}^{\prime 2} \,
    \big] \; \delta (\vec{r} - \vec{r}')
   \nn \\
&   \qquad
   +  \delta (\vec{r} - \vec{r}') \;
      \big[ 3 \, ( \vsigma_1 \cdot \vec{k} ) \, ( \vsigma_2 \cdot \vec{k} )
            -  ( \vsigma_1 \cdot \vsigma_2) \, \vec{k}^{2}
      \Big]
    \Big\}
   \nn \\
&   + \tfrac{1}{2} t_o \,
     \Big\{\Big[
       3 \, ( \vsigma_1 \cdot \vec{k}' ) \, \delta (\vec{r} - \vec{r}')\,
            ( \vsigma_2 \cdot \vec{k} )
               \nn \\
&   \qquad \qquad
            -  ( \vsigma_1 \cdot \vsigma_2 ) \, \vec{k}' \cdot \,
        \delta (\vec{r} - \vec{r}') \, \vec{k}
        \Big]
   \nn \\
&   \qquad
    + \Big[ 3 \, ( \vsigma_2 \cdot \vec{k}' ) \, \delta (\vec{r} - \vec{r}')\,
            ( \vsigma_1 \cdot \vec{k} )
               \nn \\
&   \qquad \qquad
            -  ( \vsigma_1 \cdot \vsigma_2 ) \, \vec{k} \cdot \,
        \delta (\vec{r} - \vec{r}') \, \vec{k}'
     \Big]\Big\}
\, .
\end{align}
A few further comments on the structure of the Skyrme force
\begin{itemize}
\item
The terms multiplied by $t_0$, $t_1$ and $t_e$ act in even partial
waves, whereas the $t_2$, $W_0$ and $t_o$ terms act in odd partial waves.
\item
The $( 1 + x_0 \hat{P}_\sigma )$ factors give different strength
to the spin singlet \mbox{$S=0$} and triplet \mbox{$S=1$} channels of
the two-body interaction~\cite{Nilsson}.

\item
By construction, the spin-orbit and tensor forces act in the \mbox{$S=1$}
channel only \cite{Nilsson}; hence, multiplying them with a spin exchange
operator is redundant.

\item
Most standard parametrizations comprise only one density-dependent term,
but the \texttt{Ev8} code can handle two such terms with different
exponents as used for example in the parametrizations of
Refs.~\cite{cochet04a,lesinski06a}.
\end{itemize}
\subsubsection{Isospin representation of the EDF}
Limiting ourselves to the terms that are non-zero under the time-reversal and
parity-invariance assumed here, the central and spin-orbit interactions
of Eqs.~\eqref{eq:Skyrme:central} and \eqref{eq:Skyrme:LS}, respectively,
yield the energy functional
\begin{eqnarray}
\label{eq:skyrme:energy}
\lefteqn{
\mathcal{E}^{\text{central}} + \mathcal{E}^{\text{LS}}
} \nn \\
& = & \sum_{t=0,1}
      \Big(
               A^\rho_t [\rho_0] \, \rho_t^2
             + A^{\Delta \rho}_t \, \rho_t \, \Delta \rho_t
             + A^\tau_t          \, \rho_t \, \tau_t
      \nn \\
&   &
             - A^{T}_t \sum_{\mu, \nu} J_{t, \mu \nu} \, J_{t, \mu \nu}
             + A^{\nabla \cdot J}_t  \, \rho_t \nabla \cdot \vec{J}_t
      \Big)
\, ,
\end{eqnarray}
where $\mu$ and $\nu$ run over the spatial components $x$,
$y$ and $z$, and where $t$ runs over the isoscalar and isovector terms.
The use of the coupling constant $A^{T}_t$ with a relative minus
sign might appear as an unnecessary complication of notation, but it is
kept for consistency with the definition of the full Skyrme EDF that also
contains the so-called time-odd terms \cite{per04a,hellemans12a}. Then, this term
and a time-odd one have the same coupling constant because of gauge
invariance of the functional. The coupling constants are given by
\begin{eqnarray}
\label{eq:cpl:SF}
\label{eq:func:A}
A_0^{\rho} [\rho_0]
& = & \tfrac{3}{8} t_0
       + \tfrac{3}{48} \big[   t_{3a} \, \rho_0^{\alpha_a} (\vec{r})
                             + t_{3b} \, \rho_0^{\alpha_b} (\vec{r})
                       \big]
      \nn \, ,\\
A_1^{\rho} [\rho_0]
& = & - \tfrac{1}{4}  t_0 \big( \tfrac{1}{2} + x_0 \big)
      \nn \\
&   &
      - \tfrac{1}{24} \big[  t_{3a} \big( \tfrac{1}{2} + x_{3a} \big) \,
                             \rho_0^{\alpha_a} (\vec{r})
                           + t_{3b} \big( \tfrac{1}{2} + x_{3b} \big) \,
                             \rho_0^{\alpha_b} (\vec{r})
                      \big]
     \nn \, , \\
A_0^{\tau}
& = &   \tfrac{3}{16} \, t_1
      + \tfrac{1}{4} t_2 \; \big( \tfrac{5}{4} + x_2 \big)
      \nn\, , \\
A_1^{\tau}
& = & - \tfrac{1}{8} t_1 \big(\tfrac{1}{2} + x_1 \big)
      + \tfrac{1}{8} t_2 \big(\tfrac{1}{2} + x_2 \big)
      \nn\, , \\
A_0^{\Delta \rho}
& = & - \tfrac{9}{64} t_1
      + \tfrac{1}{16} t_2 \big( \tfrac{5}{4} + x_2 \big)
      \nn \\
A_1^{\Delta \rho}
& = &    \tfrac{3}{32} t_1 \big( \tfrac{1}{2} + x_1 \big)
       + \tfrac{1}{32} t_2 \big( \tfrac{1}{2} + x_2 \big)
      \nn\, , \\
A_0^{T}
& = & - \tfrac{1}{8} t_1 \big( \tfrac{1}{2} - x_1 \big) \,
      + \tfrac{1}{8} t_2 \big( \tfrac{1}{2} + x_2 \big)
      \nn \, ,\\
A_1^{T}
& = & - \tfrac{1}{16} t_1
      + \tfrac{1}{16} t_2
      \nn \, ,\\
A_0^{\nabla J}
& = & - \tfrac{3}{4} W_0
      \nn\, , \\
A_1^{\nabla J}
& = & - \tfrac{1}{4} W_0 \, .
\end{eqnarray}
Not all of these are independent, though. The two $A_t^{\nabla J}$ are
proportional to each other, and also the two $A^T_t$ can be expressed through the
$A^\tau_t$ and $A^{\Delta \rho}_t$, cf.\ Table~I of Ref.~\cite{erler11a}.

The contribution of the tensor force (\ref{eq:Skyrme:tensor}) to the EDF
is given by \cite{per04a,ben09a}
\begin{equation}
\label{eq:EF:tensor}
\mathcal{E}^{\text{tensor}}
  =   \sum_{t=0,1}
      \Big(
      - B^{T}_t \sum_{\mu, \nu} J_{t, \mu \nu} \, J_{t, \mu \nu}
       -\tfrac{1}{2} \, B^{F}_t
      \sum_{\mu, \nu} J_{t, \mu \nu} J_{t, \nu \mu}
      \Big)
     \, ,
\end{equation}
where the summation over $\mu$ and $\nu$ is again over spatial components
$x$, $y$ and $z$. The tensor force
(\ref{eq:Skyrme:tensor}) yields two bilinear combinations
of the spin-current tensor density $J_{t, \nu \mu}$, one
symmetric and the other one asymmetric. A third combination,
$\big( \sum_{\mu} J_{\mu \mu} \big)^2$, is
only non-zero when parity is not conserved and hence vanishes in \texttt{Ev8}.
Again, the coupling constants $B^{T}_t$ and $B^{F}_t$, their signs and
the factor $1/2$ are defined to be consistent with the complete functional
including also time-odd terms~\cite{per04a,hellemans12a}.
They are related to the parameters of the Skyrme force
(\ref{eq:EF:tensor}) by \cite{per04a}
\begin{subequations}
\begin{alignat}{4}
\label{eq:func:B}
B^{T}_0
& =   - \tfrac{1}{8} (t_e + 3 t_o) \, ,
      & \qquad
B^{T}_1
& =   \phantom{-} \tfrac{1}{8} (t_e - t_o) \, ,
      \\
B^{F}_0
& =   \phantom{-} \tfrac{3}{8} (t_e + 3 t_o) \, ,
      & \qquad
B^{F}_1
& =   - \tfrac{3}{8} (t_e - t_o)
\, .
\end{alignat}
\end{subequations}
Obviously, the two isoscalar coupling constants are proportional to each other,
as are the isovector coupling constants.
%
%
\subsubsection{Combining central, spin-orbit and tensor interaction}
\label{sect:EDF:skyrme:vc+vt}
Combining Eqs.~(\ref{eq:skyrme:energy}) and ~(\ref{eq:EF:tensor}),
the complete expression for the Skyrme energy density representing
central (\ref{eq:Skyrme:central}), spin-orbit (\ref{eq:Skyrme:LS}),
and tensor (\ref{eq:Skyrme:tensor}) interactions is given by
\begin{eqnarray}
\label{eq:EF:full}
\mathcal{E}_{\text{Sk}}
& = &   \mathcal{E}^{\text{central}}
      + \mathcal{E}^{\text{tensor}}
      + \mathcal{E}^{\text{LS}}
      \nn \\
& = & \sum_{t=0,1}
      \bigg\{  C^\rho_t [\rho_0] \, \rho_t^2
             + C^{\Delta \rho}_t \, \rho_t \,  \Delta \rho_t
             + C^\tau_t          \, \rho_t \, \tau_t
      \nn \\
&   & \qquad
             - C^{T}_t \sum_{\mu, \nu} J_{t, \mu \nu} \, J_{t, \mu \nu}
             - \tfrac{1}{2} C^{F}_t
                   \sum_{\mu, \nu} J_{t, \mu \nu} \, J_{t, \nu \mu}
      \nn \\
&   & \qquad
             + C^{\nabla \cdot J}_t  \rho_t \nabla \cdot \vec{J}_t
      \bigg\}
\, .
\end{eqnarray}
It can be shown that this expression contains all possible bilinear
terms up to second order in derivatives that can be constructed from
local densities and that are invariant under spatial and time inversion,
rotations, and gauge transformations \cite{per04a}.

The expressions for the terms that are bilinear in the spin-current
tensor density, which are usually dubbed "tensor terms"
\cite{ben09a,les07a}, can be recoupled in terms of the
vector and spherical pseudotensor densities of Eq.~\eqref{eq:J:spherical}
\begin{eqnarray}
\label{eq:tensor:spherical}
\lefteqn{
- C^{T}_t \sum_{\mu, \nu} J_{t, \mu \nu} \, J_{t, \mu \nu}
- \tfrac{1}{2} C^{F}_t \sum_{\mu, \nu} J_{t, \mu \nu} \, J_{t, \nu \mu}
} \nn \\
& = & C^{J1}_t \vec{J}_t^2
      + C^{J2}_t \sum_{\mu, \nu} J^{(2)}_{t, \mu \nu} \, J^{(2)}_{t, \mu \nu}
      \, ,
\end{eqnarray}
with coupling constants given by \cite{ben09a}
\begin{eqnarray}
C^{J1}_t
& = & -\tfrac{1}{2} C^T_t + \tfrac{1}{4} C^F_t \, ,  \\
C^{J2}_t
& = & - C^T_t - \tfrac{1}{2} C^F_t \, ,
\end{eqnarray}
and where we drop a term bilinear in the pseudoscalar density.
Only the term bilinear in the vector spin-current density $\vec{J}_t^2$
in (\ref{eq:tensor:spherical}) contributes at spherical shape,
such that the two different Cartesian tensor terms cannot be distinguished
there. For further discussion of the tensor terms we refer to
Refs.~\cite{hellemans12a,ben09a}.

In a framework based on density-dependent forces the coefficients
of the energy density are related to those of the interactions
(\ref{eq:Skyrme:central}), (\ref{eq:Skyrme:LS}), and (\ref{eq:Skyrme:tensor})
through the relations
\begin{eqnarray}
\label{eq:func:C}
C_t^{\rho}[\rho_0]
& = & A_t^{\rho}[\rho_0] \, , \nn \\
C_t^{\tau}
& = & A_t^{\tau} \, , \nn \\
C_t^{\Delta\rho}
& = & A_t^{\Delta\rho} \, , \nn \\
C_t^{\nabla J}
& = & A_t^{\nabla J} \, , \nn \\
C_t^{T}
& = & A_t^{T} + B_t^{T} \, , \nn \\
C_t^{F}
& = & B_t^{F} \, ,
\end{eqnarray}
for \mbox{$t=0$}, 1. Both central and tensor interactions contribute
to the terms proportional to $C_t^{T}$.

\subsubsection{Proton-neutron representation of the functional}
\label{sect:EDF:skyrme:func:pn:representation}

While the isospin representation of the EDF in Eq.~\eqref{eq:EF:full}
is more appropriate to analyze its physics content,
a representation explicitly using proton and neutron densities
is more convenient for numerical implementations.
Then, the Skyrme energy density is written in the form
\cite{hellemans12a,bonche87}
\begin{eqnarray}
\label{eq:func:Skyrme:pn}
\mathcal{E}_{\text{Sk}}
& = &   \mathcal{E}^{\text{central}}
      + \mathcal{E}^{\text{tensor}}
      + \mathcal{E}^{\text{LS}}
      \nn \\
& = &          b_1 \; \rho^2
             + b_3 \; \rho \, \tau
             + b_5 \; \rho \, \Delta \rho
     \nn \\
&   &
             + b_{7a} \; \rho^{2 + \alpha_a}
             + b_{7b} \; \rho^{2 + \alpha_b}
             + b_{9}  \; \rho \vnabla \cdot \vec{J}
     \nn \\
&   &
             + b_{14} \sum_{\mu, \nu} J_{\mu \nu} \, J_{\mu \nu}
             + b_{16} \sum_{\mu, \nu} J_{\mu \nu} \, J_{\nu \mu}
       \nn \\
&   &  + \sum_{q=n,p} \Big(
               b_2 \; \rho^2_q
             + b_4 \; \rho_q \, \tau_q
             + b_6 \; \rho_q \, \Delta \rho_q
     \nn \\
&   &
             + b_{8a} \; \rho^{\alpha_a} \, \rho^2_q
             + b_{8b} \; \rho^{\alpha_b} \, \rho^2_q
             + b_{9q} \;  \rho_q \vnabla \cdot \vec{J}_q
      \nn \\
&   &
             + b_{15} \sum_{\mu, \nu} J_{q, \mu \nu} \, J_{q, \mu \nu}
             + b_{17} \sum_{\mu, \nu} J_{q, \mu \nu} \, J_{q, \nu \mu}
             \Big)
\, ,
\end{eqnarray}
where densities without isospin index are total densities
$\rho (\vec{r}) \equiv \rho_p  (\vec{r}) + \rho_n (\vec{r})$ and
similar for the others.\footnote{
Even though the `total' local densities are identical to the isoscalar
local densities as defined for example through Eq.~\eqref{eq:rho:iso}, we
use a different notation to clearly distinguish between the isospin
representation and the proton-neutron representation used in our codes.
}
This choice is not unique, and several other conventions
coexist in the literature \cite{erler11a,sadoudi13a}.
The coupling constants are given by
\begin{eqnarray}
\label{eq:func:Skyrme:b}
b_{1}   & = & \tfrac{1}{2}  t_0    ( 1 + \tfrac{x_0}{2} )
              \, , \nn \\
b_{2}   & = &-\tfrac{1}{2}  t_0    ( \tfrac{1}{2} + x_0 )
              \, , \nn \\
b_{3}   & = & \tfrac{1}{4} [ t_1 ( 1 + \tfrac{x_1}{2} )
                            +t_2 ( 1 + \tfrac{x_2}{2} )]
              \, , \nn \\
b_{4}   & = & -\tfrac{1}{4} [ t_1 ( \tfrac{1}{2} + x_1 )
                             -t_2 ( \tfrac{1}{2} + x_2 )]
              \, , \nn \\
b_{5}   & = & -\tfrac{1}{16} [ 3 t_1 (1 + \tfrac{x_1}{2})
                              -  t_2 (1 + \tfrac{x_2}{2})]
              \, , \nn \\
b_{6}   & = & \tfrac{1}{16} [ 3 t_1 ( \tfrac{1}{2} + x_1 )
                             +  t_2 ( \tfrac{1}{2} + x_2 ) ]
              \, , \nn \\
b_{7a}  & = & \tfrac{1}{12} t_{3a} ( 1 + \tfrac{x_{3a}}{2} )
              \, , \nn \\
b_{8a}  & = &-\tfrac{1}{12} t_{3a} ( \tfrac{1}{2} + x_{3a} )
              \, , \nn \\
b_{7b}  & = & \tfrac{1}{12} t_{3b} ( 1 + \tfrac{x_{3b}}{2} )
              \, , \nn \\
b_{8b}  & = &-\tfrac{1}{12} t_{3b} ( \tfrac{1}{2} + x_{3b} )
              \, , \nn \\
b_{9}   & = & -\tfrac{1}{2}  W_0
              \, , \nn \\
b_{9q}  & = & -\tfrac{1}{2}  W_0
              \, , \nn \\
b_{14}
        & = &            - \tfrac{1}{8} ( t_1 x_1 + t_2 x_2 )
              + \tfrac{1}{4} ( t_e + t_o )
              \, , \nn \\
b_{15}  & = &               \tfrac{1}{8} ( t_1 - t_2 )
               - \tfrac{1}{4} ( t_e - t_o )
              \, , \nn \\
b_{16}
        & = & - \tfrac{3}{8} ( t_e + t_o )
              \, , \nn \\
b_{17}  & = & \tfrac{3}{8} ( t_e - t_o )
              \, .
\end{eqnarray}
%
%
\subsubsection{Practical implementations of the Skyrme energy functional}
\label{sect:EDF:skyrme:func}
An impressive number of Skyrme parametrizations has been published since
the beginning of the 70's. They fall into two categories: some have been
constructed for a Skyrme force (Eqs.~\eqref{eq:Skyrme:central} to
\eqref{eq:Skyrme:tensor}), the others for a functional
(Eq.~\eqref{eq:EF:full}), without linking the coefficients $C_i$ to
the $t_i$, $x_i$ and $W_0$.
The differences between these groups of EDFs become
much more obvious when time-reversal invariance is broken.
Then, many additional terms appear in the functional
\cite{hellemans12a,per04a,dob96d}.
\begin{itemize}
\item
\emph{Full equivalence between the EDF and a Skyrme force.} \\
In this case, there is a strict relation between the coefficients
of the functional and the Skyrme force parameters.\footnote{It
has to be stressed that also in this case only the particle-hole
part of the EDF is derived from a density-dependent Skyrme force,
not its pairing part. The parametrizations SkP \cite{dob84} and SkS1-SkS4
\cite{gomex92a} have been fitted to provide also the pairing energy,
but this option is rarely used and not provided by \texttt{Ev8}.
Also, it is not guaranteed that a parametrization that has been
constructed as a density-dependent Skyrme force can be consistently
used as such, as many of these parametrizations exhibit finite-size
instabilities in one or the other spin-isospin
channel \cite{hellemans12a,lesinski06a,pot10a,schunck10b,hellemans13a},
sometimes even nuclear matter instabilities signaled by Landau parameters
\cite{cao10a}, at densities found in finite nuclei.}
The relations \eqref{eq:cpl:SF} to \eqref{eq:func:C} and
\eqref{eq:func:Skyrme:b} between the coefficients of the density-dependent Skyrme
force and the EDF are all verified. This means also that there is
only one parameter for the spin-orbit interaction
($A_0^{\nabla J}= 3 A_1^{\nabla J}$ or $b_9=b_{9q}$). In practice,
there are only a few parametrizations that fall into that category.
Let us quote in particular the Lyon parametrizations SLy5, SLy7
\cite{Cha98a}, the T$IJ$s \cite{les07a} and the BSk's \cite{chamel08a}.
\item
\emph{Parametrizations adjusted to the data as an EDF.} \\
The coefficients of the EDF are determined without reference
to an underlying force. In this case, all coupling constants of
the functional can be treated as being independent. For the present
case, this means that the six coupling constants $b_9$, $b_{9q}$,
$b_{14}$, $b_{15}$, $b_{16}$, and $b_{17}$ for the spin-orbit and tensor
terms can be chosen independently from the others. For a Skyrme
force this can only be done for three among them, one for the spin-orbit
terms and two for the tensor terms. However, as can be seen from
Eq.~\eqref{eq:func:Skyrme:b}, there is a one-to-one correspondence
between the coefficients of the central part of the Skyrme
interaction and the coefficients $b_1$ to $b_{8b}$ of the EDF.
This means that for the time-even part of the Skyrme functional,
the parametrization can be given equivalently in terms of the
coefficients $t_i$, $x_i, i=1$ to $3$ of the central part of the
EDF, which is common practice in the literature.
Some examples of parametrizations falling into this category
are the recent UNEDF parametrizations \cite{unedf1,unedf2}.
\item
\emph{Hybrid parametrizations.} \\
This category groups parametrizations that were adjusted as
if they were derived from a density-dependent force, but with
the coefficients of one or a few terms of the EDF modified in such
a way that the link of the coefficients to a force is broken for
some terms, but not for all of them.
For example, in many of the early Skyrme parametrizations, the
contribution of the central part of the interaction to the tensor
terms is set to zero, mainly for computational reasons.
A popular example of such a hybrid is SLy4 \cite{Cha98a}.
\end{itemize}
As said above, the differences between these variants become much
more obvious when time-reversal invariance is broken. In that case,
parametrizations fitted as Skyrme forces are sometimes used as hybrids,
or hybrids used as general functionals.


\subsection{The Coulomb energy}

The direct term of the Coulomb energy is given by
\begin{equation}
\label{Ecoul}
E_{\rm Coul}^d
= \frac{e^2}{2}
  \iint \! d^3r \, d^3r^\prime \;
  \frac{\rho_p(\vec{r}) \, \rho_p(\vec{r^\prime})}
       {|\vec{r}-\vec{r^\prime|}}
\, ,
\end{equation}
where $\rho_p(\vec{r})$ is the proton density that is assumed to be the
charge density.
One of the spatial integrations can be eliminated through calculation of
the Coulomb potential of the protons in the nucleus, which
obeys the electrostatic Poisson equation
\begin{equation}
\label{PoissonEq}
\Delta U (\vec{r})
= - 4 \pi e^2 \rho_p(\vec{r})\, ,
\end{equation}
where $e^2 = 1.43996446 \, \text{MeV} \, \text{fm}$ is the square of the
elementary charge, using the latest NIST-recommended value \cite{Mohr10}.
When solving this equation, boundary conditions need to be imposed at the edge of
the box. To that end, the Coulomb potential is expanded on the spherical
harmonics outside the boundary of the mesh and is approximated to the terms up to
\mbox{$\ell=2$}
\begin{equation}
\label{eq:BoundaryConditions}
U(\vec{r})
= \frac{e^2 Z}{r}
  + e^2 \frac{\langle \hat{Q}_{20} \rangle Y_{20}(\vec{r}) + \langle \hat{Q}_{22} \rangle \, \Re Y_{22}(\vec{r})}{r^3}\, ,
\end{equation}
where the multipole moments $Q_{20}$, $Q_{22}$ are defined in terms of the spherical
harmonics $Y_{20}$, $Y_{22}$ as in Eq.~(\ref{DefMultipole}).

The direct Coulomb energy is then calculated as
\begin{equation}
\label{Ecoul:2}
E_{\rm Coul}^d
= \int \! d^3r \; U(\vec{r}) \, \rho_p(\vec{r}) \, .
\end{equation}
Instead of its numerically costly exact calculation, the Coulomb exchange energy
is calculated in the very efficient local Slater approximation \cite{bender03a}
\begin{equation}
\label{eq:Slater}
E_{\rm Coul}^{e}
= - \frac{3 e^2}{4} \left( \frac{3}{\pi} \right)^{1/3}
  \int \! d^3r \; \rho^{4/3}_p (\vec{r}) \, ,
\end{equation}
that was used during the fit of almost all widely-used Skyrme
parametrizations and which provides in general a satisfying
approximation \cite{anguiano01a,lebloas11a}.

%
%

\subsection{Centre-of-mass correction}

The breaking of translational invariance by approximating a finite
self-bound system with a localized independent-particle state leads
to an admixture of states with finite momentum to the nuclear ground
state. Their contribution to the total binding energy can be estimated
to be
\begin{equation}
\label{eq:Ecm:full}
E_{\text{cm}}
= - \sum_{q = n,p} \frac{1}{2M} \,
  \langle \hat{\vec{P}}^2_q \rangle
= E_{\text{cm},1} + E_{\text{cm},2} \, ,
\end{equation}
where $\hat{\vec{P}}$ is the total momentum operator,
\mbox{$\hat{\vec{P}} = \sum_{k \gtrless 0} \hat{\vec{p}}_k$} with
$\hat{\vec{p}} \equiv -\textrm{i} \hbar \vnabla_k$, and
\mbox{$M = N m_n + Z m_p$} the total mass of the nucleus.
This so-called centre-of-mass correction can be separated
into a one-body

\begin{equation}
\label{eq:Ecm:1b}
E_{\text{cm},1}
= - \sum_{q = n,p} \frac{\hbar^2}{2M} \,
  2 \sum_{k > 0} v_k^2 \, \langle \Psi_k | \vnabla^2 | \Psi_k \rangle
\end{equation}
and a two-body part
\begin{equation}
\label{eq:Ecm:2b}
E_{\text{cm},2}
= - \sum_{q = n,p} \frac{\hbar^2}{2M} \,
  \sum_{k,m > 0} v_k v_m \, ( v_k v_m + u_k u_m ) \,
  \Big( |\vnabla_{km}|^2 + |\vnabla_{k\bar{m}}|^2 \Big)
\, ,
\end{equation}
where the $\vnabla_{km} = \langle \Psi_k | \vnabla | \Psi_m \rangle$ are single-particle matrix
elements of the nabla operator. Note that the nabla operator has
only matrix elements between single-particle states of opposite parity.

For numerical reasons, most parametrizations of the Skyrme EDF take into account
the one-body part only, which amounts to a renormalization of the kinetic energy
\begin{equation}
E_{\text{kin}} - E_{\text{cm},1}
= \sum_{q = n,p} \frac{\hbar^2}{2 m_q} \,
  \bigg( 1 - \frac{m_q}{Nm_n + Zm_p} \bigg) \,
  \int \! d^3r \; \tau_q (\vec{r}) \, .
\end{equation}
It is usually included in the variational equations and contributes to the total
energy and to the single-particle Hamiltonian of Eq.~\eqref{eq:hf:h}.
When, however, the two-body contribution is also taken into account, the total
centre-of-mass correction is reduced to about a third of the one-body
contribution~\cite{ben00b}.
Among the few parametrizations which include the two-body part
in the variation of the energy, one can quote SLy6 and SLy7 \cite{Cha98a}.
There exist also some parametrizations, such as SkI3, SkI4~\cite{reinhard95a},
and SV-min \cite{klu09a}, for which
the one-plus-two-body correction is only calculated
\textit{a posteriori}, and contributes perturbatively to the
total energy, but not to the single-particle energies.
Finally, let us note that for some parametrizations no
centre-of-mass correction is introduced at all. This choice is motivated by the ambiguity
of its definition when studying fusion or fission. Examples are
SLy4d (a refit of SLy4 for use in TDHF calculations~\cite{kim97a}), UNEDF1
\cite{unedf1}, and UNEDF2 \cite{unedf2}.
The \texttt{Ev8} code can handle all of these choices.

For some parametrizations, such as SkX~\cite{bro98}, the centre-of-mass correction
is approximated by a simple analytical $A$-dependent formula. This is not calculated by the
code, but it can be trivially subtracted from the total binding energy when the code is
run without centre-of-mass correction.

%
%
\subsection{Pairing correlations}
\label{sect:pairing}

In addition to a HF-like mode, where pairing correlations are neglected and the
lowest single-particle levels are occupied at each iteration, the code can handle
several choices for the pairing interaction. The first two of these have been
widely used in early applications of the HF+BCS method.
\begin{enumerate}
\item \textit{Monopole / constant strength pairing} \\
$\bar{v}^{\text{pair}}_{k \bar{k} m \bar{m}} \equiv -G_q$, with $G_q>0$.
For each nucleon species \mbox{$q=n$}, $p$, the pairing gap
(\ref{eq:Delta}) is a constant independent of the single-particle
state \mbox{$\Delta_{k \bar{k}} \equiv \Delta_q$} . The pairing
energy is then given by \mbox{$E_{\text{pair},q} = - \Delta_q^2 / G_q$}.
Such a form for pairing should definitely not be used for loosely-bound
nuclei \cite{dob84,dob96a}.

\item \textit{Constant gap pairing} \\
Monopole pairing where the pairing gap $\Delta_q$ has a
fixed value for each nucleon species~\cite{blo76a}. The pairing strength $G_q$ can then be
calculated from Eq.~\eqref{eq:Delta} and, at convergence, this option is equivalent
to a monopole pairing calculation. However, it avoids a collapse of pairing correlations
during the iterative process and is sometimes useful to stabilize the initial stages
of the mean-field iterations during which the single-particle states can vary rapidly.
This option has also been used to generate wave functions for a generator coordinate
study of pairing vibrations~\cite{MBD91a,HVB01a}.

\item \textit{Zero-range / contact pairing interaction}\\
This interaction can depend on the nuclear density and has the form
\begin{equation}
\label{eq:v:ULB}
\hat{v}^{\text{pair}}_{q}(\vec{r},\vec{r}')
= - V_q \,
  \bigg[ 1 - \alpha \; \frac{\rho_0(\vec{R})}{\rho_s} \bigg] \,
  \delta (\vec{r}-\vec{r}') \, ,
\end{equation}
where \mbox{$\rho_s = 0.16 \, \text{fm}^{-3}$},
 $V_q>0$, and $\vec{R}$ is defined as in
Eq.~\eqref{eq:Skyrme:central}. For time-reversal invariant BCS states,
the matrix elements \cite{krieger90a} are given by
\begin{eqnarray}
\label{eq:vpair:ULB}
\bar{v}^{\text{pair}}_{q,k \bar{k} m \bar{m}}
& = & - V_q
      \int \! d^3r \;
      \bigg[ 1 - \alpha \; \frac{\rho_0(\vec{r})}{\rho_s} \bigg] \,
      \nn \\
&   & \quad \times
      \Psi^\dagger_{k} (\vec{r}) \, \Psi^\dagger_{\bar{k}} (\vec{r}) \,
      \Psi_m (\vec{r}) \, \Psi_{\bar{m}} (\vec{r})
      \, ,
\end{eqnarray}
from which $\Delta_{k\bar{k}}$ and the pairing energy are
determined through Eqs.~\eqref{eq:Delta} and~\eqref{eq:epair}. Note that
the state dependent pairing gaps $\Delta_{k\bar{k}}$
depend on the overlaps between pairs of particles.

Depending on the parameter $\alpha$, the pairing interaction
(\ref{eq:v:ULB}) will be mainly active on the surface of the nucleus
(`surface pairing') for \mbox{$\alpha = 1$} \cite{dob96a,ter95a,rig99a,ben00a}
or in the core of the nucleus (`volume pairing') for
\mbox{$\alpha = 0$} \cite{dob96a,krieger90a,ben00a}.
The so-called `mixed pairing' \cite{cwiok96,dob01a,bertsch09a}
corresponds to \mbox{$\alpha = 1/2$}.
\end{enumerate}
A cutoff of the pairing is introduced to avoid a basis-size dependence
of the total energy, which would ultimately lead to divergence for all
pairing options used here~\cite{BFH85a,dob96a,yu03a}.
The code uses either a Fermi function that cuts at
$\Delta \epsilon_q$ above the Fermi energy \cite{BFH85a}
\begin{equation}
\label{eq:pair:cutoff:1}
f_k
= \big[ 1 + \text{e}^{(\epsilon_k - \lambda_q - \Delta \epsilon_q) / \mu_q}
  \big]^{-1/2} \, ,
\end{equation}
or the product of two such Fermi functions that cut at $\Delta \epsilon_q$
above and below the Fermi energy \cite{rig99a}
\begin{equation}
\label{eq:pair:cutoff:2}
f_k
= \big[ 1 + \text{e}^{(\epsilon_k - \lambda_q - \Delta \epsilon_q)/\mu_q}
  \big]^{-1/2}
  \big[ 1 + \text{e}^{(\epsilon_k - \lambda_q + \Delta \epsilon_q)/\mu_q}
  \big]^{-1/2} \, ,
\end{equation}
where the $\Delta \epsilon_q$ for neutrons and protons are input parameters,
denoted by {\sc encut} and {\sc epcut}, respectively and $\mu_q$ is fixed to $0.5 \, \text{MeV}$.
The restriction of pairing correlations to only a few single-particle states below the Fermi energy
prevents a mechanical increase of the pairing energy when the number of neutrons or protons is increased. In the same way,
 the neutron pairing energy is artificially larger if all the states below the Fermi level are included for heavy nuclei, where the number of
neutrons is much higher than the number of protons.

As it is well known, one must be careful when using the BCS approximation for nuclei far from stability. In coordinate space representation, as soon as the pairing
window includes single-particle states beyond the continuum threshold,
these begin to form an nonphysical particle gas \cite{dob84,dob96a}. Such a problem
is artificially hidden when using a basis confining the nucleons in space.

Finally, the dispersion of the particle-number defined as
\begin{equation}
\label{eq:PartDisp}
\langle(\Delta \hat{N}_q)^2\rangle
\equiv \langle\hat{N}^2_q\rangle - \langle \hat{N}_q \rangle^2
= 4 \sum_{k > 0} u_k^2 v_k^2
\end{equation}
provides a measure of pairing correlations.

\subsubsection{BCS+LN scheme}

The Lipkin Nogami (LN) prescription is used to
enforce pairing correlations in the weak-pairing limit, where
a BCS scheme would break down to the trivial HF solution without pairing.
This option of the code is meaningless when using constant gap pairing.

The LN method is an approximation to a variation after particle-number
projection. It is not fully variational, however. The variation of the
BCS Routhian~\eqref{eq:BCS:2} is replaced by the variation of
\begin{equation}
\label{eq:BCSLN}
\frac{\delta}{\delta v_j}
\Big[
2 \sum_{k > 0} \epsilon_k v^2_k
- \lambda_q \langle \hat{N}_q \rangle
+ E_{\text{pair}}
- \lambda_{2,q} \langle (\Delta \hat{N_q})^2 \rangle
\Big]
= 0 \, ,
\end{equation}
where the parameter $\lambda_{2,q}$ is not a Lagrange multiplier, but calculated
from the expression~\cite{quentin90a}
\begin{equation}
\lambda_{2,q}
= \frac{\langle{\hat{v}^{\text{pair}}_{q}} \,
         (\Delta \hat{N}_q^2-\langle\Delta \hat{N}_q^2\rangle)\rangle \,
         \langle\Delta \hat{N}_q^2\rangle
       -\langle{\hat{v}^{\text{pair}}_{q}} \, \Delta \hat{N}_q\rangle \,
        \langle\Delta \hat{N}_q^2\rangle}
       {(\langle\Delta \hat{N}_q^4\rangle
        - \langle\Delta \hat{N}_q^2\rangle^2) \,
         \langle\Delta \hat{N}_q^2\rangle
        - \langle\Delta \hat{N}_q^3\rangle^2} \, ,
\end{equation}
whose dependence on $v_k$ is ignored during the variation.
The occupation numbers of the BCS state \eqref{eq:BCS}
are obtained as~\cite{ben00a}
\begin{equation}
v_k^2
= \frac{1}{2}
  \left[ 1
        - \frac{\epsilon^\prime_k - \lambda_q}
               {\sqrt{ (\epsilon^\prime_k - \lambda_q)^2 + f^2_k \, \Delta^2_{k \bar{k}}}}
  \right]\, ,
\end{equation}
and where
\begin{equation}
\label{eq:epsac}
\epsilon^\prime_k
= \epsilon_k + 4 \lambda_{2,q} \,(v^2_k- 0.5) \, ,
\end{equation}
with $\epsilon_k$ the expectation values of the single-particle
Hamiltonian \eqref{eq:hf:equation}. The effect of the LN correction
on the single-particle spectrum can be seen in this equation:
the levels far below the Fermi level ($v^2_k\simeq 1$) are pushed up and those far above are pulled down, rendering the spectrum more compressed and favouring the presence of pairing correlations.
However, one should not forget that the LN prescription is an approximation
of a particle-number projection and is thus going beyond a simple quasi-particle
model. Therefore, the meaning of single-(quasi)particle energies is not
clearly defined anymore. Note that the quasi-particle energies \eqref{eq:Equasi:BCS}
printed by \texttt{Ev8} are calculated from the energies $\epsilon'_{k}$
\eqref{eq:epsac} instead of the single-particle energies when
the LN prescription is applied.

The total binding energy is corrected for particle-number fluctuations
by adding
\begin{equation}
\label{eq:LN:correction}
E_{\text{LN}}
= -\sum_{q = p,n} \lambda_{2,q} \, \langle (\Delta \hat{N}_q)^2 \rangle
\, .
\end{equation}
One-body operators can be corrected in a simpler manner by calculating
effective LN occupation numbers \cite{bennour89a,rei96a}
\begin{equation}
w_k
= v_k^2
  + \frac{ s^2_k \; \big( \sum s^2 \big) \, \Big\{ \, c_k \sum s^2 - \sum s^2 c \, \Big\}
         }
         {2 \big( \sum s^2 \big) \,
           \Big\{     \big( \sum s^2 \big)^2
                  - 3 \sum s^4
                  + 2 \sum s^2
           \Big\}
          - 4 \, \big( \sum s^2 c \big)^2
         }
\, ,
\end{equation}
where we use the shorthands $s_i \equiv \big| 2 u_i v_i \big|$,
$c_i \equiv v_i^2 - u_i^2$, as well as $\sum s^2 \equiv \sum_{i > 0} s_i^2$
and similar for other sums.
Corrected expectation values of time-even hermitian one-body operators
can then be calculated as in \cite{bennour89a,rei96a}
\begin{equation}
\langle \hat{O} \rangle_{\text{LN}}
= 2 \sum_{k > 0} w_k \, O_{kk}\, ,
\end{equation}
where the $O_{kk}$ are the single-particle matrix elements of $\hat{O}$.
The \texttt{Ev8} code prints several multipole moments calculated this
way.

\section{The shape of the nuclear density}

The shape of the nuclear density distribution can be analyzed in terms of the
multipole moments
\begin{equation}\label{DefMultipole}
\langle \hat{Q}_{\ell m} \rangle
= \langle \hat{r}^{\ell} \hat{Y}_{\ell m} \rangle\, .
\end{equation}
Because of the symmetries imposed in \texttt{Ev8}, these moments are real and obey
the relation $\langle \hat{Q}_{\ell m} \rangle = \langle \hat{Q}_{\ell \, -m} \rangle$.
They vanish identically for all but even values of $\ell$ and $m$.
\texttt{Ev8} provides the values of these moments up to
\mbox{$\ell=6$} at the end of the iterative process.

\subsection{Radii}

Mean-square (ms) and root-mean-square (rms) radii for the neutron, proton, and mass densities are computed by the \texttt{Ev8} code, together with the neutron skin $\Delta r_{np}$. They are defined as
\begin{eqnarray}
 r^2_{q,\text{ms}}
& = & \frac{1}{N_q} \int \! d^3r \; r^2 \rho_q (\vec{r})\, ,\\
r_{q,\text{rms}}
& = & \sqrt{ r^2_{q,\text{ms}} }\, ,\\
\Delta r_{np}
& = & r_{n,\text{rms}} - r_{p,\text{rms}}\, ,
\end{eqnarray}
where we follow the definition of the neutron skin of Ref.~\cite{Myers69}.
To be compared to experimental data, radii have to be corrected for the
composite nature of nucleons and their extended charge
distribution \cite{bender03a}. For charge radii, a simple estimate is
provided by adding the mean-square radius of the proton to the
mean-square radius of the point proton distribution $r_{p,\text{ms}}$
printed by the code,
$r_{c,\text{ms}} = r_{p,\text{ms}} + 0.64 \, \text{fm}$ \cite{BBH08a}.

\subsection{Quadrupole Moments}
\label{QuadruSection}

The quadrupole moments play a central role in collective models~\cite{RingSchuck,Rowe70a}.
Here, they are used as constraints to generate energy surfaces. For these reasons,
the code provides information on several parametrizations of these moments.

In their Cartesian representation, they are defined as the expectation value of
\begin{alignat}{2}
\label{MultiX}
 \hat{Q}_x & =- \sqrt{\frac{4 \pi}{5}} \left(\hat{Q}_{20} - \sqrt{6} \hat{Q}_{22}\right) \;  & = &  \;  2\hat{x}^2 - \hat{y}^2 - \hat{z}^2\, , \\
 \hat{Q}_y & =- \sqrt{\frac{4 \pi}{5}} \left(\hat{Q}_{20} + \sqrt{6} \hat{Q}_{22}\right)  & = & \;   2\hat{y}^2 - \hat{x}^2 - \hat{z}^2 \, ,\\
 \hat{Q}_z & = \sqrt{\frac{ 16 \pi}{5}}  \hat{Q}_{20} & = & \;  2\hat{z}^2 - \hat{x}^2 - \hat{y}^2\, .
 \label{MultiZ}
\end{alignat}
An alternative representation is given in terms of
the deformation parameter $q$ and the triaxiality angle $\gamma$. These
parameters are related to the Cartesian quadrupole moments by

\begin{alignat}{2}
q
& = \sqrt{\frac{16 \pi}{5} \left\langle \hat{Q}_{20}^2 + \hat{Q}^2_{22} \right\rangle} \;
 =  \sqrt{\frac{2}{3} \left\langle \hat{Q}^2_{x} + \hat{Q}^2_y + \hat{Q}^2_z \right\rangle} \, ,\\
\gamma &= 2 \arctan\left( \frac{\sqrt{2}\langle \hat{Q}_{22} \rangle}{\sqrt{\langle \hat{Q}_{20}\rangle^2 + 2\langle \hat{Q}_{22} \rangle^2} +  \langle\hat{Q}_{20}\rangle}\right)
\\
& =    2 \, \text{arctan}\left( \frac{\langle \hat{Q}_x  - \hat{Q}_y \rangle}{ \sqrt{ 2 \langle \hat{Q}_z^2 - \hat{Q}_x \hat{Q}_y \rangle} + \langle \hat{Q}_x - \hat{Q}_y\rangle}  \right).
\end{alignat}

Another set of variables, $q_1$ and $q_2$, is used to define the quadrupole constraints
(see Sect.~\ref{constraintdata}). They are related to $q$ and $\gamma$ by
\begin{eqnarray}
\label{q1}
q_{1} & = & q\cos(\gamma) - \frac{1}{\sqrt{3}}q\sin(\gamma) \, , \\
\label{q2}
q_{2} & = & \frac{2}{\sqrt{3}}q\sin(\gamma) \, .
\end{eqnarray}
or
\begin{eqnarray}
q
& = & \sqrt{q_{1}^{2} + q_{2}^{2} + q_{1}q_{2}} \, ,
      \\
\gamma
& = &  2 \, \text{arctan}\left( \frac{\sqrt{3}\, q_{2}}{\sqrt{ \left( 2\, q_1 + q_2\right)^2 + 3\, q_2^2} + 2\, q_1 + q_2}\right).
\end{eqnarray}
The connection between the Cartesian, the $q$, $\gamma$ and the $q_1$, $q_2$
representations is summarized by
\begin{alignat}{2}
\label{SummaryMulti}
\langle \hat{Q}_{x} \rangle
& = -\frac{1}{2}(q_{1} - q_{2}) \;
& = & -q\cos(\gamma + 60) \, ,
            \\
\langle \hat{Q}_{y} \rangle
& = -\frac{1}{2}(q_{1} + 2q_{2})  \;
& = & -q\cos(\gamma - 60) \, ,
            \\
\langle \hat{Q}_{z} \rangle
& = +\frac{1}{2}(2q_{1} + q_{2})  \;
& = & +q\cos(\gamma) \, .
\end{alignat}
To define a quadrupole constraint, one must fix the order of the Cartesian quadrupole moments.
Any permutation of the $x$, $y$, $z$ coordinates will indeed lead to the same total energy for the nucleus but to different wave functions.  The different possibilities are illustrated in Fig.~\ref{Qgamma}.
For a calculation limited to axial shapes, the most convenient choice is to use the same symmetry axis for prolate and oblate configurations. This can be achieved by
varying $q_1$ from negative values for oblate shapes to positive values for prolate ones, $q_2$ being always zero. This choice corresponds to the $z$-axis as symmetry axis and $\gamma=180 ^\circ$ for oblate configurations and $0 ^\circ$ for prolate ones. To explore triaxiality, one can limit the calculation to $q_1$ and $q_2$ positive, the triaxiality angle $\gamma$ varying then from $0^\circ$ to $60^\circ$ and the symmetry axis from the $z$-axis for prolate shapes to the $y$-axis for oblate ones.

\begin{figure}
\centerline{ \includegraphics[scale=0.45]{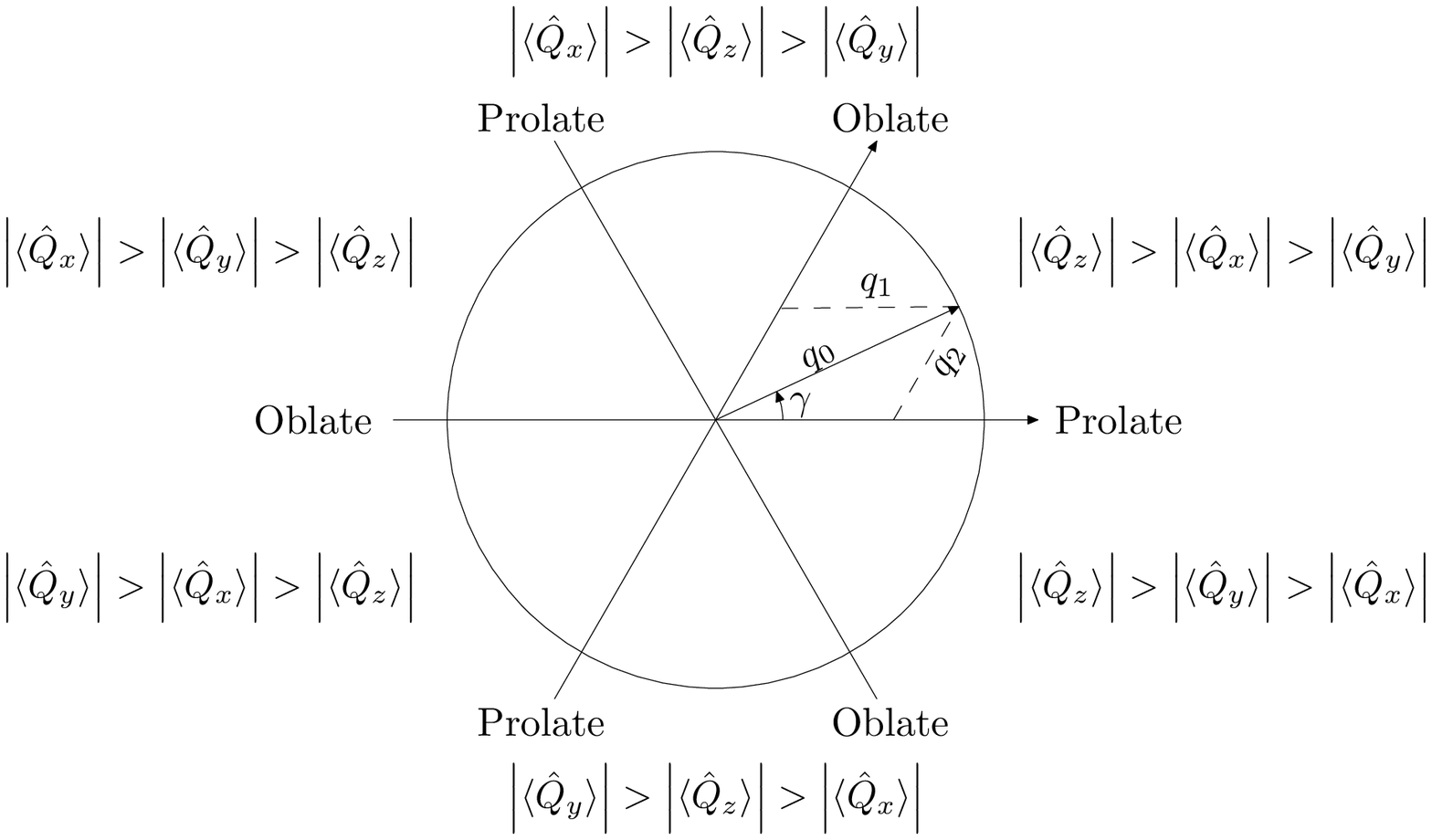}}
\caption{Graphical interpretation of the relation between the $q, \gamma$ and the $q_1, q_2$ representation of the quadrupole moments. Taking one sextant of the circle is sufficient to parametrize all possible ellipsoid shapes. Axial states are confined to the $q_1,q_2$ axes; all other states are triaxial.}\label{Qgamma}
\end{figure}

\subsection{Deformation parameters}

The deformation of a nucleus is not an observable in the strict sense.
Only for nuclei presenting large deformations in their ground state, the quadrupole moment obtained
from a mean-field model can be approximately related to $B(E2)$-values~\cite{RingSchuck,raman01a}.
Still, it is interesting to compare the deformations predicted by different models and to relate
the different ways of parameterizing the deformation.

The first way used in the code is simply a rescaling the multipole moments. We define deformation parameters $\beta_{\ell m}$ by extracting the main dependence in the total mass of the moments given by Eq.~\eqref{DefMultipole}
\begin{equation}
\label{equ:betalm}
\beta_{\ell m}
=  \frac{4 \pi}{3 R_0^\ell A} \langle \hat{Q}_{\ell m} \rangle \, ,
\end{equation}
where $R_0 = 1.2 \, A^{1/3} \, \text{fm}$ and $A$ is the mass number. When
considering deformation parameters for individual nucleon species, $A$
has to be replaced by the appropriate particle number
in Eq.~\eqref{equ:betalm}, but not in the definition of $R$.

To compare deformations between self-consistent models as used in
\texttt{Ev8} and microscopic-macroscopic models where
the mean-field potential is parametrized by a simple potential,
one can introduce deformation parameters such that the uniform
distribution of matter inside the nuclear surface defined by
the multipole expansion \cite{Hasse88}
\begin{equation}
\label{equ:nuc_surf}
R(\Omega)
= c(\alpha) \, R_0 \Big[ 1 + \sum_{\ell} \alpha_{\ell} Y_{\ell 0} (\Omega) \Big]\, ,
\end{equation}
reproduces the HF+BCS expectation value of the multipole moments.
This correspondence is analyzed by the \texttt{Ev8} code for axial
deformations with $\gamma = 0^{\circ}$ and $\gamma = 180^{\circ}$ only, and
also for quadrupole and hexadecapole moments only.
The constant $c(\alpha)$ in Eq.~(\ref{equ:nuc_surf}) is introduced
in order to conserve the same volume as a function of deformation
\cite{cwiok96,Hasse88}. Taking only the terms up
to second order in $\alpha_2$ and $\alpha_4$,
quadrupole and hexadecapole deformation parameters
$\alpha_{2}$ and $\alpha_{4}$ are solutions of the following equations
\begin{alignat}{2}
Q_{20}
&= \frac{3}{4 \pi} A R_0^{2}
   \left( \alpha_2
          + \frac{2}{7} \sqrt{\frac{5}{\pi}} \, \alpha_2^2
          + \frac{20}{77}\sqrt{\frac{5}{\pi}} \, \alpha_4^2
          + \frac{12}{7\sqrt{\pi}} \alpha_2 \, \alpha_4
   \right) \, , \\
Q_{40}
&=  \frac{3}{4 \pi} A R_0^4
    \left(   \alpha_4 + \frac{9}{7 \sqrt{\pi}} \, \alpha_2^2
           + \frac{729}{1001\sqrt{\pi}} \, \alpha_4^2
           + \frac{300}{77\sqrt{5\pi}} \, \alpha_2 \, \alpha_4
    \right) \, ,
\end{alignat}
where \mbox{$R_0 = 1.2 \, A^{1/3} \, \text{fm}$}.
Note that the values
taken by $\alpha_2$ and $\alpha_4$ are always smaller than those
of the corresponding $\beta_{\ell 0}$ given by Eq.~\eqref{equ:betalm}. Note also that $\alpha_2$ and $\alpha_4$ are sometimes referred to as  $\beta_2$ and $\beta_4$

\subsection{Constraints}

Several constraints on the mean value of operators can be imposed during the variation of
the total energy of the atomic nucleus
\begin{itemize}
\item rms radii $\sqrt{\langle \vec{\hat{r}}^2\rangle_t}$, $\sqrt{\langle \vec{\hat{r}}^2\rangle_n}$,$\sqrt{\langle \vec{\hat{r}}^2\rangle_p}$\, ,
\item $q_{1t},q_{1n}, q_{1p}$\, ,
\item $q_{2t},q_{2n}, q_{2p}$\, ,
\end{itemize}
where $q_{1,2}$ are defined as in Eqs.~\eqref{q1} and \eqref{q2} and the subscripts $t$, $p$,
and $n$ indicate that the constraints can either be placed on the total
density or on the neutron and proton density separately.

As is well known~\cite{RingSchuck,Flo73a} it is more efficient to use
quadratic constraints than linear ones. One then minimizes a modified Routhian
\begin{equation}
\label{constraint}
E_{\text{kin}} + E_{\text{ph}} + C \, \big( \langle{\hat O}\rangle -\mu \big)^2
\, ,
\end{equation}
instead of $E_{\text{kin}} + E_{\text{ph}}$ to derive the
single-particle Hamiltonian in Eq.~\eqref{eq:hf:h}. This results in
a contribution $C(\langle{\hat O}\rangle -\mu){\hat{O}}$ to the potential.
Thus, the quadratic constraint on the energy is equivalent to a linear
constraint with a Lagrange multiplier $C(\langle{\hat O}\rangle -\mu)$
that varies during the iterations.
Several constraints may be active at the same time.
At every intermediate printout, \texttt{Ev8} prints the energy associated with the
constraints and their derivatives $2 C(\langle{\hat O}\rangle -\mu)$, which represent
the slope of the associated energy surface.

The constant $C$ determines the strength of the constraint; its value should be
such that the contribution of the constraint to the energy is of the order of
a few MeV at the start of the iterations. The constant $\mu$
corresponds to the value that one wants to obtain at convergence for the matrix
element $\langle{\hat O}\rangle$. It is possible to let the code modify the value
of $\mu$ to $\mu_{\rm eff}$ during the iterations in such a way
that the constraint $\langle{\hat O}\rangle=\mu$ is satisfied with a high accuracy.
The change of $\mu$ is controlled by an input parameter {\sf epscst}.
Starting for the constraint from a value $\mu_0$ at the first iteration, the value at iteration
$i+1$ is given as a function of the value at the previous iteration by
\begin{equation}
 \mu^{(i+1)} = \mu^{(i)} - \text{{\sc epscst}} \left( \left(\langle \hat{O} \rangle\right)^{(i+1)}  - \mu_0 \right)\, ,
 \label{eq:ReadjustmentConstraints}
\end{equation}
where {\sf epscst} is a constant between 0 and 1.

The adjustment of the constraints by \texttt{Ev8} is similar to that of the augmented Lagrangian method
of Staszczak \textit{et al.}~\cite{Sta10}. The two methods mainly differ in that the parameter $\text{{\sc epscst}}$ can be
freely chosen in \texttt{Ev8}, if desired. In general, however, the
default value provided by \texttt{Ev8} leads to an excellent convergence of the constraints, as can be seen from the example in Sect. \ref{sect:AccConst}.
This procedure is used since our first calculations of super-deformed bands in the neutron deficient Hg isotopes~\cite{Gal94}

\subsection{Cutoff of the constraints}
\label{sect:CutoffMultipole}

Multipole operators are unbound for $r\rightarrow\infty$ and can lead to unphysical results
when used as constraints in mean-field calculations~\cite{Bass75,Font75} if the
wave functions are allowed to spread out very far in space.
This kind of instabilities is hidden when the wave functions are expanded on a basis that
automatically cuts them at large distances from the nucleus, but such an implicit
cutoff depends on the size
of the basis. The problem becomes more apparent in a solution of the mean-field equations on
a mesh as in \texttt{Ev8}. Therefore, the code offers the possibility to cut the constraining operators
at large distance by introducing a form factor $f(\vec{r}, \rho)$ \cite{rutz95a} in the constraints
\begin{equation}
\label{eq:Cutoff}
\langle \hat{Q}_{2m}^{\text{cut}} \rangle
= \int \! d^3r \;  f(\vec{r}, \rho) \, \rho(\vec{r}) \, Q_{2m}(\vec{r}) \, ,
\end{equation}
where $f(\vec{r}, \rho)$ is dependent on the values of
{\sf rcut, icutq} and {\sf acut}.

First, when {\sf icutq} is equal to 2, a spherical cut-off is used to calculate the quadrupole moments. It amounts to taking $f(\vec{r}, \rho)$ in Eq.~\eqref{eq:Cutoff} density independent
\begin{eqnarray}
  f(\vec{r}, \rho) & = & \left\{ \begin{array}{ll}
                      \frac{ \exp (-d)}{  1 + \exp \left( -d \right)} & \text{if d $\geq 0$} \, ,\\
                      \frac{ 1 }{ 1 + \exp \left( d \right)} & \text{if d $\leq 0$}\, ,
                     \end{array} \right.\\
d & = & \frac{|\vec{r}| -  \text{\sc{rcut}}}{\text{\sc{acut}}}\, .
  \label{eq:cutoff:2}
\end{eqnarray}
A second option is used when {\sf icutq} is equal to 1. This procedure
uses a density dependent cutoff factor
\begin{equation}
f(\vec{r}, \rho)
= \frac{\rho(\vec{r})}{\text{\sf rcut}}
  \left[ 1 - \tanh^2 \frac{\rho(\vec{r})}{\text{\sf rcut}} \right]
   + \tanh \frac{\rho(\vec{r})}{\text{\sf rcut}}\, .
    \label{eq:cutoff:1}
\end{equation}
A third, and recommended, option was first proposed in Ref.~\cite{rutz95a}.
It is used when {\sf icut} is equal to 0. The factor $f(\vec{r}, \rho)$ is
\begin{equation}
\label{eq:cutoff:0}
f(\vec{r}, \rho)
= \frac{1}{1 + \exp \left( \frac{d - \text{{\sf rcut}}}{\text{{\sf acut}}} \right)}
\, .
\end{equation}
In this expression the number $d$ is the distance between the point $\vec{r}$ and the
surface  $\rho_{\text {eq}_t} = 0.1 \max \left\{\rho_t \right\}$.

\section{Rotational Properties}

For a deformed nucleus, the single-particle wave functions are in general not
eigenstates of the angular momentum operators, such that the value of
$\langle \Psi_k | \hat{j}_{z} | \Psi_k \rangle$
is not restricted to half-integer multiples of $\hbar$. Its value is printed at each
intermediate printout, as well as the value $j_{k}$ defined in units of $\hbar$ by
\begin{equation}
 \langle \Psi_k | \hat{j}^2 | \Psi_k \rangle = j_k ( j_k + 1) .
\end{equation}
Furthermore, the symmetries assumed in \texttt{Ev8} lead to vanishing
$\langle\hat{j}_x\rangle$ and $\langle\hat{j}_y\rangle$. For the same reasons,
the expectation values of the many-body angular momentum operators $\hat{J}_{\mu}$,
$\mu = x$, $y$, $z$, are also zero. The quantities  $\langle \hat{J}_{\mu}^2 \rangle$
and $\langle \hat{J}^2 \rangle$ are in general neither zero, nor restricted to
specific multiples of $\hbar$.

In addition, an estimation of the Belyaev moment of inertia is determined
using~\cite{RingSchuck,Rowe70a}
\begin{equation}
\label{eq:belyaev}
\Theta_{\text{Belyaev},\mu}
= 2 \hbar^2 \sum_{i,j \gtrless 0}
 \frac{| \langle \Psi_i | \hat{j}_\mu | \Psi_j \rangle)|^2 (u_i v_j - v_i u_j)^2}
      {E_{\text{qp},i} + E_{\text{qp},j}}
\, ,
\end{equation}
where the $E_{\text{qp},k}$ are the quasiparticle energies of Eq.~\eqref{eq:Equasi:BCS}.
Unlike other quantities calculated by the code, the sum in Eq.~\eqref{eq:belyaev} is not
cut by the occupation numbers, but by the quasiparticle energies in the denominator.
As a consequence, some single-particle states above the pairing cutoff that contribute to Eq.~\eqref{eq:belyaev} might not
be contained in the space considered for the calculation, leading to a slight
underestimation of $\Theta_{\text{Belyaev},\mu}$.

As an alternative, the code also prints the rigid-rotor moments of inertia
\begin{equation}
\Theta_{\text{rigid},\mu}
= m \int d^3 \vec{r} \, \rho(\vec{r}) \, r_{\mu}^{\perp} \, ,
\end{equation}
for $\mu = x$, $y$, and $z$, where $\rho_{\mu}^{\perp}$ is the distance
to the axis of rotation.

In the asymmetric rotor model~\cite{Rowe70a}, the
spurious rotational energy of a deformed BCS state
can then be estimated as
\begin{equation}
E_{\text{rot}}
= \sum_{\mu} \frac{\langle \hat{J}^2_\mu \rangle}{2 \Theta_\mu}
\, ,
\end{equation}
which for an axially-symmetric nucleus can be rewritten as
$E_{\text{rot}} = \langle \hat{J}^2 \rangle/ (2 \Theta_\perp)$,
with $\Theta_\perp$ being the moment of inertia perpendicular to
the symmetry axis. Note that the matrix elements of
$\hat{J}_{\mu}^2$ are printed by the code in units of $\hbar^2$
and the moments of inertia in units of $\hbar^2 / \text{MeV}$,
such that the rotational energy in MeV can be easily constructed.

\section{Numerical algorithms}
\label{NewNumerics}

\subsection{Derivatives on a mesh}
\label{sect:Deriv}

Our calculations are performed on a 3D mesh. It implies that the wave functions are discretized and that their values at the mesh points are the variational parameters. The two parameters controlling the accuracy of the calculation are the distance $dx$
between the discretization points and the dimension of the box in which the wave functions are different from zero. We consider only meshes with equidistant points.
Integrals are then simply calculated by summing the values of the discretized functions at the mesh
points.
Note that the dimension of the box can be different along the three Cartesian axes, which is useful when studying very elongated shapes.

In the previous version of the code, derivatives were calculated using finite-difference formulae.
It is still the case during the iterations but all derivatives entering the EDF are recalculated after convergence using a more precise method.
It has indeed been shown~\cite{BH86a} that any function defined by its value on a set of
$N$ equidistant points can be expressed as a function of $N$ orthonormal functions, called Lagrange functions,

\begin{equation}\label{lag_f}
  f_r(x) = \frac{1}{N} \frac{\sin \left[\frac{\pi}{dx} (x-r \, dx)\right]}{\sin\left[\frac{\pi}{dx}\frac{x-r \, dx}{N}\right]}\, ,
\end{equation}
where $r \, dx$ are the mesh points. These functions, called Lagrange interpolation functions, have the property that they are equal to $1$ when $x=r \, dx$ and zero for  all the other mesh points. For a 3D mesh, three sets of such functions are defined along the three Cartesian axes. In the following, we limit ourselves to one direction. Centering the mesh around the origin and considering an even $N$, we obtain non-integer values of $r$ that vary from $-(N-1)/2$ to $(N-1)/2$ in steps of 1.

Then, the function $\phi(x)$ can be calculated at any point between  $\pm dx \, (N+1)/2$ by means of
\begin{equation}\label{phix}
\phi(x)
= \sum_r \phi(x_r) \, f_r(x) \, .
\end{equation}
It is equal to zero at $\pm dx \, (N+1)/2$ and takes the values $\phi(x_r)$ at the mesh points.
This expression can be used to calculate the values of the wave functions at the points of a
different mesh, defined either by a change of the mesh size or a rotation.

It also offers an alternative to the use of finite-difference formulas for the calculation
of derivatives on a mesh.
The expression for the first order Lagrangian derivative of $f_t(x)$ at a mesh point $s \, dx$ is given by
\begin{equation}
\label{der1}
\frac{d f_t(x)}{dx}\bigg|_{x=x_s}
= \left\{ \begin{array}{ll}
          {\displaystyle
          (-1)^{t-s}\frac{\pi}{Ndx} \frac{1}{\sin(\pi (t-s)/N)} } &
          \text{for $t \neq s$,} \\
          0 &
          \text{for $t = s$.}
          \end{array} \right.
\end{equation}
The second-order Lagrangian derivative is given by
\begin{equation}
\label{der2}
\frac{d^2 f_t(x)}{dx^2}\bigg|_{x=x_s}
= \left\{ \begin{array}{ll}
          {\displaystyle
           (-1)^{t-s}\left(\frac{2\pi}{Ndx}\right)^2
           \frac{\cos \left[\pi (t-s)/N\right]}{\sin^2\left[\pi (t-s)/N\right]}
           }
           & \text{for $t \neq s$}, \\
           {\displaystyle
           \left(\frac{\pi^2}{3dx}\right)^2 \left(1-\frac{1}{N^2}\right) }
           & \text{for $t = s$.}
          \end{array} \right.
\end{equation}
Hence, the calculation of the derivatives of a function at the mesh points amounts to a matrix multiplication, implying $N^2$ multiplications for each direction.
Then, the numerical cost increases rapidly with the number of mesh points. Therefore, from a numerical point of view it is more efficient
to use finite-difference formulas during the mean-field iterations. More precisely, the first order derivative is approximated by a third-order finite-difference scheme
 while a fourth-order scheme is chosen for the second-order derivative. While less precise, their numerical cost only increases linearly with the number of mesh points.
Finally, Lagrange derivatives based on Eqs.~\eqref{der1} and~\eqref{der2} are used at the end of the iterative procedure to obtain a higher accuracy. Detailed tests~\cite{RBH14} demonstrate that
this scheme significantly reduces the computing time while maintaining a high numerical accuracy on the total energy.

For the numerical integration used, the derivatives \eqref{der1}
and \eqref{der2} are numerically exact such that
expressions that differ by partial integration become numerically equivalent.
Also, applying twice the first derivative \eqref{der1} on a mesh function is
numerically equivalent to applying once the second derivative \eqref{der2} on
the same function. Neither is the case when using finite-difference formulas
for the derivatives.

%
%
%
\subsection{Numerical scheme}

The HF+BCS equations are solved iteratively using the imaginary time
step method~\cite{Dav80}. At each iteration, the code successively
executes the following steps
\begin{enumerate}
 \item Compute the mean-field densities using the occupation numbers $(v_k^{(i)})^2$. \label{step1}
 \item Advance in imaginary time:
$\left| \Phi_{k}^{(i+1)} \right\rangle =  \left( 1  - \frac{\delta t}{\hbar} \hat{h} \right) \left| \Psi_k^{(i)} \right\rangle$.
 \item Orthonormalise the single-particle wave functions \\
 $\left| \Phi^{(i+1)}_k \right\rangle \rightarrow \left|  \Psi_k^{(i+1)}\right\rangle$.
 \item Solve the BCS equations for the occupation numbers $(v_k^{(i+1)})^2$.
 \item Check for convergence. If the convergence criteria are satisfied, stop. Otherwise, return to step \ref{step1}.
\end{enumerate}

\subsubsection{Step 1: Computing the densities}
\label{sec:CompDen}
The mean-field densities are recalculated at each iteration using to the expressions given in \cite{bonche87}. However,
to use directly these new densities to construct the mean-field potentials might lead to numerical instabilities. It is usually safer to construct the new mean-field with  densities that are averaged between two iterations
\begin{eqnarray}
\label{eq:DenDamp}
\rho^{(i+1)} (\vec{r})
& = & \frac{\text{\sc nxmu}}{100}
      \sum_{k \gtrless 0} \left( v_k^{(i)} \right)^2 \, \Psi^{(i+1) \dagger}_k (\vec{r}) \,
      \Psi^{(i+1)}_k (\vec{r})
      \nn \\
&   &    + \frac{100 - \text{\sc nxmu}}{100} \, \rho^{(i)} (\vec{r}) \, ,
\end{eqnarray}
where the input parameter {\sc nxmu} is an integer between 0 and 100.
Experience has shown that a value $\textsc{nxmu}=25$ is a reasonable choice.
Similar prescriptions, using the same parameter {\sc nxmu}, are used for the
kinetic energy density $\tau (\vec{r})$ and the divergence of the spin-orbit current
$ \vnabla \cdot \textbf{J} (\vec{r})$.

The contribution of the constraints to the mean-field Hamiltonian are
also averaged over two iterations with another input parameter \textsc{ral}. The contribution to the single-particle Hamiltonian coming from a constraint on the matrix element of an operator $\hat{O}$ at iteration $i+1$, $\hat{h}^{(i+1)}_{\hat{O}}$ is given by
\begin{equation}
 \hat{h}^{(i+1)}_{\hat{O}} = \left( 1 - \textsc{ral} \right) \hat{h}^{(i)}_{\hat{O}} + \textsc{ral}\, 2 \, C \left( \langle \hat{O}\rangle - \mu^{(i)}\right) \hat{O}
\end{equation}
for a quadratic constraint, as defined in Eq. \eqref{constraint}. The case of linear constraints is analogous.

The occupation probabilities of the single-particle states used to compute
the mean-field densities are those calculated at the previous mean-field
iteration.

\subsubsection{Step 2: Advancing in imaginary time }
\label{sec:advImTime}

The number of wave functions that can be constructed on a mesh is very large. With 20 mesh points in  each direction, which is a typical number for a medium mass nucleus using a step size of $0.8 \, \text{fm}$, this number is equal to 64 000 taking into account
the symmetries imposed in the code. Most of them, however, do not have any relevance and do not need to be calculated.
To obtain a good convergence on the mean-value of the particle number operators, the BCS occupation of the most excited single-particle states that are calculated should be lower than $10^{-4}$.
Taking into account the two-fold Kramers degeneracy of the states, this implies that the number of single-particle wave functions to be explicitly included in the calculation
is of the order of the number of neutrons and of protons. The imaginary-time method~\cite{Dav80} is very well adapted to perform a diagonalization limited to a few low-lying states.
The principle
is to use the exponential of the mean-field Hamiltonian as a filter
\begin{equation}
\label{eq:ImTime0}
\left| \Phi_{k}^{(i+1)} \right\rangle
= \exp \left( - \frac{\delta t}{\hbar} \hat{h}\right)
  \left| \Psi^{(i)}_k \right\rangle \, ,
\end{equation}
where $\delta t$ is the size of the time step. Since the transformation
defined by Eq.~\eqref{eq:ImTime0} is not unitary,
the states $\Phi^{(i+1)}_{k}$ need to be orthonormalized at each iteration,
resulting in the new single-particle states $\Psi^{(i+1)}_{k}$.

In practice, the exponential operator is approximated by a Taylor expansion
to first order
\begin{equation}\label{eq:ImTime}
\left| \Phi_{k}^{(i+1)} \right\rangle
=  \left( 1  - \frac{\delta t}{\hbar} \hat{h} \right)
  \left| \Psi^{(i)}_k \right\rangle \, .
\end{equation}
At each iteration and for each $\Psi^{(i)}_{k}$, the weight of the eigenstates of $\hat{h}$ corresponding to eigenvalues larger than $\epsilon_k$ must be decreased.
This is only possible if $\frac{\delta t}{\hbar}$ times the largest eigenvalue of $\hat{h}$ is lower than one. As such, it fixes an upper limit for $\delta t$ which could otherwise be chosen
as large as possible to obtain a fast convergence. An upper bound of the largest eigenvalue that can be obtained on a mesh is provided by the largest possible kinetic energy.
The latter is obtained for a single-particle wave function constant in absolute value but changing sign at each mesh point. This upper limit of the kinetic energy increases
when the accuracy of derivatives on a mesh is increased and is the largest when using the derivatives of Eq.~\eqref{der2}. As such, the use of finite-difference formulas during the iterative process permits a larger choice for the value of $\delta t$.
As a reference, for a mesh spacing $dx=0.8 \, \text{fm}$, $\delta t = 0.012 \times 10^{-22} \, \text{s}$ is a safe choice. For a spacing of $dx=0.64 \, \text{fm}$, the time step has to be decreased, a value of $\delta t= 0.01 \times 10^{-22} \, \text{s}$ being usually sufficient. Note that these values depend slightly on the parametrization; for parametrizations with very small effective masses, one may have to decrease $\delta t$.

\subsubsection{Step 3: Orthonormalising the single-particle wave functions}

As mentioned above, the imaginary-time method generates a set of states $\Phi_k^{(i+1)}$ that are not orthonormal.
The well-known Gram-Schmidt procedure is applied to determine the states $\Psi_k^{(i+1)}$ from the $\Phi_k^{(i+1)}$.

At each iteration, the diagonal matrix elements of the
single-particle Hamiltonian are calculated
\begin{equation}
\epsilon^{(i+1)}_k
= \langle \Psi_k^{(i)} | \hat{h} | \Psi_k^{(i)} \rangle \, ,
\end{equation}
and are used to solve the BCS equations. It is only at convergence
that the $\epsilon_k$ become eigenvalues of $\hat{h}$.

\subsubsection{Step 4: Solve the BCS equations}

The single-particle energies and wave functions obtained in step 2 are used to solve the pairing equations of Sect.~\ref{sect:pairing}. The $(v^{(i)})^2_k$ are determined iteratively. The iterations are halted when the Fermi energy varies by less than $1 \, \text{keV}$. This ensures that the mean number of particles is obtained with an accuracy higher than $10^{-5}$ particles.

For each single-particle state, the occupations ($v^{(i)})^2_k$, the pairing gaps $\Delta^{(i)}_{k\overline{k}}$ and the quasiparticle energies are printed at every intermediate printout.

\subsubsection{Step 5: Check of convergence}
\label{sec:ConverCheck}

There are several ways to check the convergence of a mean-field calculation.
The code terminates the iteration process automatically when four criteria are met, with tolerances that are fixed in the data. To avoid an accidental stop of the calculation at a given iteration, these criteria must be met during 7 successive iterations. They concern:

\begin{itemize}
 \item The relative change in total energy $E$
 \begin{equation}
 \left|\frac{E^{(i)} - E^{(i-j)}}{E^{(i)}}\right|  \leq \text{{\sc epse}}  \quad j = 1,2,\ldots,7.
 \end{equation}
 \item The change in Fermi energy for both particle species
 \begin{equation}
 \left|\lambda_{pn}^{(i)} - \lambda_{pn}^{(i-j)}\right|  \leq \text{{\sc epsf}}  \quad j = 1,2,\ldots,7.
 \end{equation}
 \item When it is constrained, the change of quadrupole moment. The relative difference from the required value must be lower than a given tolerance. The condition must be satisfied for the three Cartesian components of the quadrupole moment
 \begin{equation}
 \left|\frac{\langle \hat{Q}_{x,y,z} \rangle -  Q_{x,y,z}}{Q_{x,y,z}}\right|  \leq \text{{\sc epsq}}\, ,
 \end{equation}
 where $Q_{x,y,z}$ is the desired value.

 In case of a spherical configuration, the meaning of EPSQ is different and the absolute value of each component is checked
 \begin{equation}
 \left|\langle \hat{Q}_{x,y,z} \rangle \right |  \leq \text{{\sc epsq}}\, .
 \end{equation}
 \item The sum of the dispersions of the single particle energies, weighted by their occupation probabilities
\begin{equation}
 \left| \frac{2}{A} \sum_{k > 0} v^2_k \left [ \langle \Psi_k | \hat{h}^2 | \Psi_k\rangle -  \langle \Psi_k | \hat{h} | \Psi_k\rangle^2\right] \right| \leq\text{{\sc epsdh}}\, .
\end{equation}
 \end{itemize}

\subsection{Accuracy of the calculations}

For a given EDF (mean and pairing fields), the code provides a numerically approximate solution of the mean-field problem. Instead of a number of oscillator shells and of the frequencies of the oscillator, the parameters governing the numerical accuracy are:

\begin{itemize}
  \item The box size should be large enough to avoid an artificial
cut of the tail of the single-particle wave functions.
  \item The mesh spacing $dx$ that governs the accuracy of the calculation of derivatives.
  \item The Coulomb boundary conditions that are not exact.
\end{itemize}

\subsubsection{Box Size \& Discretization of the mesh}

An extensive test of the accuracy of our numerical scheme will be published in a separate paper \cite{RBH14}, where we will in particular show that our method is very well suited to calculate nuclei for a very large range of deformations. Here, we extract a few results from this study that can guide the choice of the number of mesh points and the size of the box required to obtain a given accuracy. In Table~\ref{tab:BoxSizeTable}, we illustrate the effect of the size of the box on the energy of the three spherical nuclei $^{40}$Ca, $^{132}$Sn and $^{208}$Pb. The mesh size used in the calculation is $1.0 \, \text{fm}$.
In terms of the number of points, a convergence better than $1 \, \text{keV}$ is achieved at 12 points for $^{40}$Ca, at 14 for $^{132}$Sn and at 18 for $^{208}$Pb.
For smaller mesh sizes, the same physical dimension of the box should be used.

In Table~\ref{tab:BoxParamTable}, we confront the energies obtained with \texttt{Ev8} to those calculated with the spherical HFB code \texttt{Lenteur} \cite{Karim} for the same nuclei. Because of
the large number of discretization points in the radial coordinate, the latter values can be considered exact. In all cases, a mesh size of around $0.6 \, \text{fm}$ results in an accuracy on the total energy of a few keV. In general, for a mesh size of $0.8 \, \text{fm}$, the accuracy is better than $100 \, \text{keV}$, which is sufficient for most applications.  The accuracy of the previous version of the code is presented in the last column of Table~\ref{tab:BoxParamTable} . It is less good because of the first-order discretization for the Laplacian in the Coulomb solver and because of the finite-difference formulae for derivatives. The current improvements made in the code increase the accuracy by more than an order of magnitude. In the past, the smaller accuracy of the old version of \texttt{Ev8} was circumvented by recalculating the energies with the Lagrange derivatives in a separate code.

\begin{table}[h!]
 \centering
 \begin{tabular}{|l|l|l|}
  \hline
  $nx$ & $E$ (MeV)  \\
  \hline
10  &$-344.26415$ \\
12  &$-344.13149$ \\
14  &$-344.13171$ \\
16  &$-344.13176$ \\
18  &$-344.13178$ \\
20  &$-344.13179$ \\
  \hline
 \end{tabular}
 \begin{tabular}{|l|l|l|}
  \hline
  $nx$ & $E$ (MeV)  \\
  \hline
  10  & $-1102.7113$ \\
  12  & $-1102.9050$ \\
  14  & $-1102.9144$ \\
  16  & $-1102.9150$ \\
  18  & $-1102.9152$ \\
  20  & $-1102.9152$ \\
  \hline
  \end{tabular}
  \begin{tabular}{|l|l|l|}
  \hline
  $nx$ & $E$ (MeV)  \\
  \hline
16  & $-1634.8269$ \\
18  & $-1634.8274$ \\
19  & $-1634.8275$ \\
20  & $-1634.8277$ \\
21  & $-1634.8278$ \\
22  & $-1634.8278$ \\
\hline
\end{tabular}
\caption{
Energy of $^{40}$Ca (top left), $^{132}$Sn (top right) and $^{208}$Pb (bottom)
as a function of box size, using the SLy4 parametrization without pairing and a step size of $1.0 \, \text{fm}$. The Lagrange formulas have been used to calculate derivatives.}
\label{tab:BoxSizeTable}
\end{table}
\begin{table}
\begin{tabular}{|l|l|l|l|l|}
 \hline
 $nx$ & $dx$ (fm) & $E$ & $\Delta E$  & $\Delta E$ (v1) \\
 \hline
 13 & 1.00 & $-344.1317$  & 0.1   & 2    \\
 17 & 0.76 & $-344.2499$  & 0.01  & 0.2  \\
 21 & 0.62 & $-344.2612$  & 0.003 & 0.03 \\
 25 & 0.52 & $-344.2631$  & 0.001 & 0.07 \\
 29 & 0.45 & $-344.2638$  & 0.0005& 0.07 \\
 33 & 0.39 & $-344.2642$  & 0.0001& 0.06 \\
 \hline
\end{tabular}
 \begin{tabular}{|l|l|l|l|l|}
 \hline
 $nx$ & $dx$ (fm) & $E$  & $\Delta E$  & $\Delta E$ (v1) \\
 \hline
 18 & 1.0  & $-1103.4233$ & 0.1   & 3  \\
 20 & 0.77 & $-1103.5115$ & 0.05  & 1  \\
 22 & 0.70 & $-1103.5395$ & 0.02  & 0.8\\
 24 & 0.64 & $-1103.5496$ & 0.01  & 0.3\\
 26 & 0.59 & $-1103.5538$ & 0.005 & 0.1\\
 28 & 0.55 & $-1103.5558$ & 0.003 & 0.02\\
 \hline
 \end{tabular}
\begin{tabular}{|l|l|l|l|l|}
 \hline
 $nx$ & $dx$ (fm) & $E$  & $\Delta E$  & $\Delta E$ (v1) \\
 \hline
 20  & 1.0  & $-1634.8277$ & 0.9   & 14  \\
 24  & 0.83 & $-1635.5735$ & 0.1   & 4   \\
 28  & 0.71 & $-1635.6751$ & 0.03  & 1   \\
 32  & 0.63 & $-1635.6945$ & 0.01  & 0.3 \\
 36  & 0.55 & $-1635.6997$ & 0.003 & 0.07\\
 \hline
\end{tabular}
\caption{Energy of $^{40}$Ca (top), $^{132}$Sn (middle) and $^{208}$Pb (bottom) for different box parameters, using the SLy4 parametrization without pairing and calculated with Lagrange derivatives. Parameters have been chosen to give the same volume for the box in each case. The error $\Delta E$ is obtained as a difference between our results and those obtained with the spherical code \textsc{Lenteur} (see text). The difference with the energies  obtained with the previous version (v1) of \texttt{Ev8} are given in the last column. All energies are in MeV.
}
\label{tab:BoxParamTable}
\end{table}

\subsubsection{Precision of the Coulomb Solver}
The precision of the Coulomb solution is determined by several factors, in particular: the dimensions of the mesh, the order of the discretisation of the Laplacian and the boundary conditions.

The boundary conditions are determined using an expansion of the Coulomb potential limited to $\ell=0$ and $\ell=2$ as can be seen in Eq.~\eqref{eq:BoundaryConditions}.
This may be too limited for very elongated shapes. For such configurations, the accuracy of the calculation can be verified by
adding points to the box in which the Poisson equation is solved (parameters {\sc npx, npy} and {\sc npz}, see Sect. \ref{sect:data}) and, if necessary, increase this number in the mean-field calculation. In this way, the contribution of $\ell \geq 4$ terms to the boundary conditions is reduced.

Two options are available to discretize the Laplacian to solve the Poisson equation. The second order discretization ({\sc icoul}=2) should be systematically used, the first order approximation is kept for compatibility reasons with the previous version of \texttt{Ev8}.

\subsubsection{Total energy from the single-particle energies}

In self consistent mean-field methods~\cite{RingSchuck}, the total energy can be calculated in two different ways,
either directly from the EDF or from the single-particle energies. When
calculated from the single-particle energies $\epsilon_k$, the total energy is given as
\begin{equation}
\label{eq:sum:espe}
E
= \sum_{k>0} v_k^2  \epsilon_k  + \frac{E_{\text{kin}}}{2} + E_{\text{SR}}
  + \frac{1}{3} \, E^{e}_{\text{Coul}} + E_{\text{C}} + E'_{\text{corr}} + E_{\text{pair}} \, ,
\end{equation}
where $E_{\text{kin}}, E^{e}_{\text{Coul}}$ and $ E_{\text{pair}}$ are
defined as in Sect.~\ref{sect:EDF}, where $E_{\text{C}}$ is the contribution of the
constraints to the single-particle
energies that has to be removed from the calculation of the total energy
\begin{equation}
\label{eq:EfromSp}
E_{\text{C}}
= \sum_{\text{constrained } \hat{O}} C_{\hat{O}} \left(  \langle \hat{O} \rangle  -  \mu_{\hat{O}} \right)\langle \hat{O} \rangle \, ,
\end{equation}
and where $\frac{1}{3} E^{e}_{\text{Coul}}$ and
\begin{equation}
E_{\text{SR}}
  =   - \frac{1}{2} \int \! d^3r \, \sum_{x = a,b}  \alpha_x \, \rho^{\alpha_x} \,
      \bigg( b_{7x} \, \rho^2  + b_{8x} \sum_{q=n,p} \rho_q^{2} \bigg) \, ,
\end{equation}
are the rearrangement energies related to the density-dependent terms in the
Slater approximation to the Coulomb exchange energy and in the Skyrme interaction,
respectively.

The correction for spurious motion $E'_{\text{corr}}$ to be used in \eqref{eq:sum:espe},
however, might be slightly different
from $E_{\text{corr}}$ introduced in Sect.~\ref{sect:EDF}. Depending on the options chosen for the code
(more specifically the parameter {\sf ncm2}, see Sect.~\ref{sect:data}), part of the
centre-of-mass correction is already included in the single-particle energies $\epsilon_{k}$.
Similarly, the LN correction $E_{\text{LN}}$, Eq.~\eqref{eq:LN:correction}, has to be added
for {\sf npair}$=3$ and 5.
Thus we define $E'_{\text{corr}}$ as the parts of $E_{\text{corr}}$ that are not included
in $\sum_{k>0} v_k^2 \, \epsilon_k$.

 Theoretically, the equality of these two methods for calculating the energy is obtained at convergence. In practice however, the numerical approximations made in the calculation do not have the same effect on both terms and the equality is only approximate at convergence. This is due in particular to the approximate treatment of the derivatives on the mesh and to the fact that the single-particle wave functions are set to zero at the border of the mesh instead of at infinity. We summarize in Table~\ref{tab:EfuncvsEsp}
how close one can expect the two energies to be as a function of the mesh size. A larger difference indicates that the box size is most probably not large enough and the single-particle wave functions are artificially set to zero at too small a distance.

\begin{table}[t]
\centering
 \begin{tabular}{|l|l|l|}
  \hline
  $nx$ & $dx$ (fm) & $\Delta E$ (MeV)\\
  \hline
  20 & 1.00 & 0.321 \\
  24 & 0.83 & 0.113 \\
  28 & 0.71 & 0.047 \\
  32 & 0.62 & 0.022 \\
  36 & 0.55 & 0.011 \\
  40 & 0.50 & 0.006 \\
  44 & 0.45 & 0.003 \\
  \hline
 \end{tabular}
\caption{Difference between different ways of calculating the total energy as a function of mesh size for $^{208}$Pb, using the SLy4 parametrization without pairing. Note that these numbers are calculated with finite difference derivatives.}
\label{tab:EfuncvsEsp}
\end{table}

\subsubsection{Accuracy of constrained calculations}
\label{sect:AccConst}
Figure \ref{fig:Zr84Surf} gives an example of calculation of a triaxial quadrupolar energy map. The input data are those given for  the sample run of $^{84}$Zr included in the code package. The step size for the quadrupole moments $q_1$ and $q_2$ is $100 \, \text{fm}^2$. The quadrupole moments at the end of the iterations differ from the targeted values by around $10^{-3} \, \text{fm}^2$, and the energies are converged within at least a few keV. These results demonstrates the power of our method for calculations with multiple constraints.

If the same calculations are run at smaller mesh parameter $dx=0.64 \, \text{fm}$ as compared to $dx=0.8 \, \text{fm}$ in
{Fig.~\ref{fig:Zr84Surf}, the resulting energy surface is extremely similar: the finer mesh calculation is systematically between 40 and $60 \, \text{keV}$ lower in energy than the coarse calculation.

\begin{figure}
\includegraphics{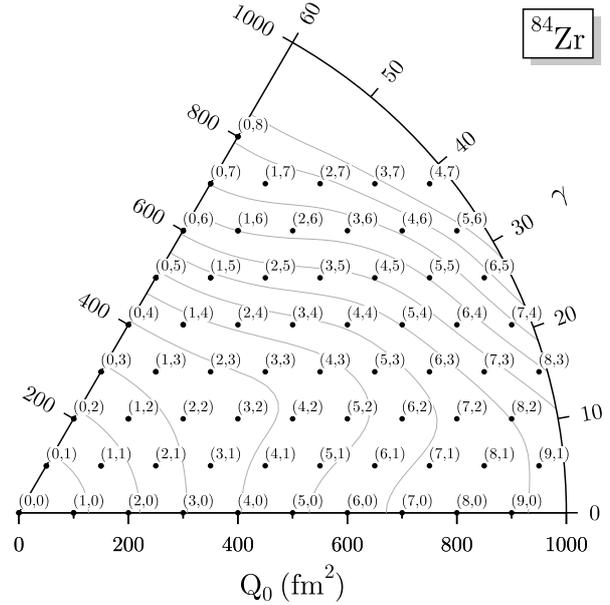}
\caption{\label{fig:Zr84Surf}
Deformation energy surface in the $(Q_{0},\gamma)$ plane for $^{84}$Zr. The dots mark the calculated points at their $(q_1, q_2)$ coordinates in units of $100 \, \text{fm}^2$. The contours are iso-energy lines, at every $1 \, \text{MeV}$ relative to the spherical ground state. Parameters are the ones distributed in the example included in the code-package, in particular, the SLy4 Skyrme parametrization and the density-dependent pairing interaction.
}
\end{figure}

\section{Comparison with other HF+BCS solvers}

While a detailed study of the accuracy of the present numerical scheme will be presented in a forthcoming paper, we now succinctly discuss its
performance with respect to methods based on an oscillator-basis expansion.
Because the equivalence between spherical (\texttt{HOSPHE}~\cite{HOSPHE1}), axial (\texttt{HFBTHO}~\cite{HFBTHO1}) and triaxial (\texttt{HFODD}~\cite{HFODD8}) oscillator-based codes was recently demonstrated by Stoitsov \textit{et al.}~\cite{Stoitsov12}, we restrict ourselves to a comparison between \texttt{Ev8} and \texttt{HOSPHE}.

In Table \ref{tab:HOSPHEComp}, we present the different components of the EDF (see Eq.~\eqref{eq:EF:schematic}) for the converged \texttt{HOSPHE} and \texttt{Ev8} results with the SLy4 parametrization. For \texttt{HOSPHE}, 50 harmonic-oscillator shells and an oscillator length of $b = 2.2345 \, \text{fm}$ were chosen. For \texttt{Ev8}, the results shown are those obtained at the smallest mesh size in Table \ref{tab:BoxParamTable}.
While an excellent agreement between the two codes is obtained for the total energy, it results from cancellations between the different components of the EDF.  These individual components can differ up to a few MeV for a heavy nucleus as $^{208}$Pb, whereas the total energy agrees up to a few keV.
This is to be expected because the total energy is the only variational quantity.
The large differences for the kinetic and the Skyrme EDF energies can be understood from (a) the different numerical treatment of the derivatives in both schemes (b) the different asymptotic behaviour of the oscillator-basis and mesh-discretized wave functions.}

 Also, it is instructive to compare the energies obtained for the ``standard'' run parameters of the codes. The results reported
in Table \ref{tab:HOSPHECompRegular} were obtained for 16 oscillator shells and the oscillator lengths suggested by \texttt{HOSPHE} \cite{HOSPHE1}. For \texttt{Ev8}, we chose the parameters given in Table~\ref{tab:BoxParamTable} for a mesh size $dx \approx 0.8 \, \text{fm}$. Then, the differences between both codes are much larger and of the order 1/1000 of the total energy. When comparing with the total energies reported in Tables~\ref{tab:HOSPHEComp} and \ref{tab:HOSPHECompRegular}, one can conclude that the accuracy obtained with \texttt{Ev8} is significantly better than that with \texttt{HOSPHE} under standard conditions.

As a final note, we remark that the agreement for the total energy between both codes and between different discretisations for the same code are the result of large cancellations in specific parts of the total energy. As example, the kinetic energy of $^{208}$Pb differs by $12 \, \text{MeV}$ between different mesh parameters, while the total energy differs by less than $200 \, \text{keV}$.

\begin{table}
\begin{tabular}{|l|r|r|r|}
\hline
 & \texttt{Ev8} & \texttt{HOSPHE} & Difference\\
\hline
$E_{\text{tot}}$ & $-344.264$ & $-344.262$ & 0.002 \\
$E_{\text{kin}}$ &  635.063  & 635.036   & 0.027  \\
$E_{\text{Skyrme}}$  & $-1051.458$ & $-1051.428$ & 0.030 \\
$E_{\text{Coul, dir}}$  & 79.620 &  79.619  & 0.001 \\
$E_{\text{Coul, exc}}$ & $-7.490$ &  $-7.490$  & 0.000 \\
\hline
\hline
 &  \texttt{Ev8} & \texttt{HOSPHE} & Difference\\
\hline
$E_{\text{tot}}$ & $-1103.556$ & $-1103.553$ & 0.003 \\
$E_{\text{kin}}$ &  2447.411  & 2446.551   & 0.860  \\
$E_{\text{Skyrme}}$  & $-3892.450$ & $-3891.536$ & 0.914 \\
$E_{\text{Coul, dir}}$ & 360.315 & 360.260 & 0.055 \\
$E_{\text{Coul, exc}}$  & $-18.831$ & $-18.829$  & 0.002 \\
\hline
\hline
 &  \texttt{Ev8} & \texttt{HOSPHE} & Difference\\
\hline
$E_{\text{tot}}$ & $-1635.700$ & $-1635.692$ & 0.007 \\
$E_{\text{kin}}$ &  3868.841  & 3866.176   & 2.665  \\
$E_{\text{Skyrme}}$  & $-6301.388$ & $-6298.483$ & 2.905 \\
$E_{\text{Coul, dir}}$ & 828.131 &  827.883 & 0.248 \\
$E_{\text{Coul, exc}}$ & $-31.278$ & $-31.269$  & 0.009 \\
\hline
\end{tabular}
\caption{Comparison of different \texttt{Ev8} quantities with \textsc{HOSPHE}, using the `best' parameters, see text. $^{40}$Ca (top) $^{132}$Sn (middle) and $^{208}$Pb (bottom). Note that the kinetic energy reported here also contains the one-body c.m.\ correction. All energies are expressed in MeV.}
\label{tab:HOSPHEComp}

\begin{tabular}{|l|r|r|r|}
\hline
 & \texttt{Ev8} & \texttt{HOSPHE} & Difference\\
\hline
$E_{\text{tot}}$ & $-344.250$ & $-344.250$ & 0.000 \\ 
$E_{\text{kin}}$ &  636.108 &  634.958 & 1.150 \\  
$E_{\text{Skyrme}}$  & $-1052.556$ & $-1051.335$  & 1.221 \\
$E_{\text{Coul, dir}}$ & 79.695  & 79.615 & 0.080 \\  
$E_{\text{Coul, exc}}$ & $-7.496$ & $-7.489$ & 0.007 \\
\hline
\hline
 &  \texttt{Ev8} & \texttt{HOSPHE} & Difference\\
\hline
$E_{\text{tot}}$ & $-1103.512$    & $-1102.934$  & 0.578\\
$E_{\text{kin}}$ & 2452.824     & 2444.220   & 8.604\\ 
$E_{\text{Skyrme}}$   & $-3898.153$ & $-3888.503$ & 9.650\\
$E_{\text{Coul, dir}}$ & 360.666  & 360.169   & 0.497\\
$E_{\text{Coul, exc}}$ & $-18.849$   & $ -18.820 $& 0.029\\
\hline
\hline
 &  \texttt{Ev8} & \texttt{HOSPHE} & Difference\\
\hline
$E_{\text{tot}}$    & $-1635.574$ & $-1634.453$ & 1.121  \\ 
$E_{\text{kin}}$    & 3880.647  & 3861.663  & 18.984 \\ 
$E_{\text{Skyrme}}$  & $-6314.142$ & $-6292.549$ & 21.593 \\
$E_{\text{Coul, dir}}$ & 829.241   & 827.687   & 1.554 \\ 
$E_{\text{Coul, exc}}$ & $-31.319$   & $-31.254$   & 0.065 \\ 
\hline
\end{tabular}
\caption{Comparison of different \texttt{Ev8} quantities with \texttt{HOSPHE} using the `usual' parameters, see text. $^{40}$Ca (top) $^{132}$Sn (middle) and $^{208}$Pb (bottom). Note that the kinetic energy reported here also contains the one-body c.m.\ correction. All energies are expressed in MeV.}

\label{tab:HOSPHECompRegular}

\end{table}


\section{Files needed to compile and run the code}
\label{sect:operation}
In this section, we succinctly describe how to run the code.

To compile the code, the user should provide
a file named \textsc{param8.h}. This file contains the compilation parameters and in particular
the number of points in each direction of the grid ({\sf mx}, {\sf my} and {\sf mz}).
Along with the parameter {\sf h} read on fort.12, they fix the dimensions of the box.
The number of single-particle wave functions (neutrons and protons combined) that will be followed during the iterations
({\sf mw}) is also specified in this file, together with the total number of points in the box \textsf{mv}, and the parameter \textsf{mq} giving the total number of variational parameters per single-particle wave function, taking into account that it has four components. Finally, it includes also the parameter {\sf mc} that can be used to calculate the Coulomb
potential in a larger box with up to {\sf mc} extra points along
each direction. The structure of {\sf param8.h} is:
\begin{verbatim}
       parameter (mx=16,my=16,mz=16,mc=10)
       parameter (mv=mx*my*mz,mq=4*mv,mw=62)
\end{verbatim}
This is a F77 code file and the data
must be entered starting column 7.

 A file named \textsf{fort.12} must be supplied to execute the code. It contains a full set of single-particle wave functions that will be the starting point of the iterations. This file can be a file generated by previous \texttt{Ev8} runs (see below) or the output of the auxiliary program \texttt{Nil8}. Note that the number of wave functions read on this file must not exceed the compilation parameter \textsf{mw} and the box parameters \textsf{mx, my, mz} should also be smaller than the compilation parameters. Smaller values are however allowed.
In addition to this starting point, some running parameters must be specified on STDIN, as explained in Sect. \ref{sect:data}.

During the calculation all diagnostic data and information are written to STDOUT. In addition, the code produces a \textsf{fort.13} which, among other information, contains the single-particle wave functions obtained at the end of the iterative process.

\section{Description of the data}
\label{sect:data}

We now describe the runtime parameters required to run of \texttt{Ev8}. As mentioned before, these parameters are read by \texttt{Ev8} from STDIN. The data are split into several blocks according to their physical meaning and they are summarized in Table \ref{tab:Input}. In the following, we provide detailed information on each of these data blocks.

\begin{table}[ht!]
\centering
\begin{tabular}{|l|l|}
\hline
 \textbf{Parameter} & \textbf{Format} \\
 \hline
 \multicolumn{2}{|c|}{\textbf{ Global information}}\\
 \hline
  {\sf head}                       &  20a4   \\
  {\sf npx,npy,npz,iCoul}          &  4i5    \\
  {\sf dt}                         &  e15.8 \\
  {\sf nitert,nxmu,ndiag }         &  3i5    \\
  {\sf nprint }                    &  1i5    \\
  {\sf npn,npp}                    &  2i5    \\
 \hline

  \multicolumn{2}{|c|}{\textbf{ Skyrme Interaction}}\\
 \hline
  {\sf afor}   &  a4     \\
 { if afor.eq.`XXXX'} & \\
  {\sf \qquad ncm2,nfunc,nmass,ncoex}     & 4i5    \\
  {\sf \qquad t0,x0 }                     & 2e15.8 \\
  {\sf \qquad t1,x1,t2,x2 }               & 4e15.8 \\
  {\sf \qquad t3a,x3a,yt3a }              & 3e15.8 \\
  {\sf \qquad t3b,x3b,yt3b  }             & 3e15.8 \\
  {\sf \qquad if (nfunc.eq.0) }	          &        \\
  {\sf \qquad\qquad wso }                 & 1e15.8 \\
  {\sf \qquad\qquad te,to  }              & 4e15.8 \\
  {\sf \qquad if (nfunc.eq.1)}            &        \\
  {\sf \qquad \qquad wso,wsoq }           & 2e15.8 \\
  {\sf \qquad\qquad b14,b15,b16,b17  }    & 4e15.8 \\
  {\sf \qquad if (nmass.eq.2)}            &        \\
  {\sf \qquad\qquad hbm(1),hbm(2)}        & 2e15.8 \\
  \hline
   \multicolumn{2}{|c|}{\textbf{Pairing Interaction}}\\
  \hline
  {\sf npair,icut}                    &  2i5    \\
  {\sf gn,encut,delmax(1),alpha}      &  4e15.8 \\
  {\sf gp,epcut,delmax(2)}            &  4e15.8 \\
  \hline
  \multicolumn{2}{|c|}{\textbf{Convergence Tresholds}}\\
  \hline
  {\sf epse,epsdh,epsq,epsf}      &   4e15.8 \\
  \hline
  \multicolumn{2}{|c|}{\textbf{ Shape Constraints}}\\
  \hline
  {\sf imtd,ifrt,imtg,icutq}          &  4i5    \\
  {\sf ral,epscst,rcut }              &  3e15.8 \\
  {\sf cqr,rtcst or cqr,rncst,rpcst}  &  3e15.8 \\
  {\sf cq2,delq}                      &  2e15.8 \\
  {\sf q1t,q2t or q1n,q2n,q1p,q2p }   &  4e15.8 \\
  \hline
  \end{tabular}
  \caption{Content of the input file, split into five distinct input blocks.}
  \label{tab:Input}
\end{table}

\subsection{Global information on the calculation}

\begin{table}[!h]
\centering
\begin{tabular}{|l|l|}
 \hline
 \multicolumn{2}{|c|}{\textbf{ Global information}}\\
 \hline
  {\sf head}                       &  20a4   \\
  {\sf npx,npy,npz,iCoul}          &  4i5    \\
  {\sf dt}                         &  e15.8 \\
  {\sf nitert,nxmu,ndiag }         &  3i5    \\
  {\sf nprint, iverb }             &  2i5    \\
 {\sf npn,npp}                     &  2i5    \\
 \hline
 \end{tabular}
 \end{table}
\begin{itemize}
\item  {\sf head}: A title to be written in the output file. Any alphanumeric field of at most 80 characters is allowed.
\item  {\sf npx, npy, npz}: The number of mesh points added to the 3D box to improve
the accuracy on the Coulomb field and Coulomb energy.
These numbers must be smaller than or equal to the value of {\sf mc} given in
the {\sf param8.h} file.
\item  {\sf iCoul}: If {\sf iCoul} is equal to 0 or 2, a five-point formula for the Laplacian is used along each Cartesian direction to solve the Poisson equation giving the direct term of the Coulomb potential. If {\sf iCoul} is equal to 1, a three-point formula is used as in the previous version of the code.
\item  {\sf dt}: Size of the imaginary time step $\delta t$ (see Eq.~\eqref{eq:ImTime}) in units of 10$^{-22}$~s. Its value must be close to the largest one allowed by the mesh size {\sf dx} read on {\sf fort.12}. See Sect. \ref{sec:advImTime}.
\item  {\sf nitert}: the number of iterations is equal to
{\sf abs(nitert)}.
       If a negative value of {\sf nitert} is given, the iteration
index is reset to zero at the start of the iterations.
       Otherwise, it starts from the number of iterations stored
on the file {\sf fort.12}.
\item  {\sf nxmu}: The mean-field densities $\rho_q (\vec{r})$,
$\tau_q (\vec{r})$ and $\nabla \cdot \vec{J_q (\vec{r})}$ of protons
and neutrons are averaged during the iterations
according to Eq.~(\ref{eq:DenDamp}).
       If {\sf nxmu} is read to be 0, it is set to 25.
\item  {\sf ndiag}:  If this parameter is equal to 1, a
diagonalization of the HF Hamiltonian in the single-particle subspace
is performed before the first HF iteration.
\item  {\sf nprint}:  Number of iterations between two
consecutive full printouts.
\item  {\sf npn, npp}: Neutron and proton number of the nucleus. Because of the
Kramers degeneracy of the wave functions,
       {\sf npn} should be less than {\sf 2*nwaven} and {\sf npp} less than
       {\sf 2*nwavep}, where {\sf nwaven} and {\sf nwavep} are the
       numbers of wave functions stored on {\sf fort.12}.
\end{itemize}

\subsection{Skyrme mean-field interaction}
\begin{table}[h!]
\centering
\begin{tabular}{|l|l|}
\hline
 \multicolumn{2}{|c|}{\textbf{ Skyrme Interaction}}\\
 \hline
 {\sf afor}   &  a4     \\
 { if afor.eq.`XXXX'} & \\
  {\sf \qquad ncm2,nfunc,nmass,ncoex}     & 4i5    \\
  {\sf \qquad t0,x0 }                     & 2e15.8 \\
  {\sf \qquad t1,x1,t2,x2 }               & 4e15.8 \\
  {\sf \qquad t3a,x3a,yt3a }              & 3e15.8 \\
  {\sf \qquad t3b,x3b,yt3b  }             & 3e15.8 \\
  {\sf \qquad if (nfunc.eq.0) }	          &        \\
  {\sf \qquad\qquad wso }                 & 1e15.8 \\
  {\sf \qquad\qquad te,to  }              & 4e15.8 \\
  {\sf \qquad if (nfunc.eq.1)}            &        \\
  {\sf \qquad \qquad wso,wsoq }           & 2e15.8 \\
  {\sf \qquad\qquad b14,b15,b16,b17  }    & 4e15.8 \\
  {\sf \qquad if (nmass.eq.2)}            &        \\
  {\sf \qquad\qquad hbm(1),hbm(2)}        & 2e15.8 \\
  \hline
\end{tabular}
\end{table}

\begin{itemize}
\item {\sf afor}: The parametrization of the Skyrme interaction may be chosen from a
      predefined set, presented in \ref{sec:PreDef}.
      To select any one of them, one must
enter {\sf afor} accordingly, e.g.\ {\sf Skm*} or {\sf SIII}, etc.
These characters are left-justified.
      To use a custom parametrization, it is sufficient to set {\sf afor} to any string which is not in the
codes' database. The Skyrme parameters below should then be entered accordingly.
\item {\sf ncm2, nfunc, nmass, ncoex }:
In contrast with previous versions of the code, these parameters do not have to be provided for predefined parametrizations. For those, they are set automatically to the values used in the fit of these parametrizations. To modify these predefined values, the parametrization has to be introduced as a new one.

{\sf ncm2}: Centre-of-mass correction
\begin{enumerate}
\addtocounter{enumi}{-3}
\item : One- and two-body c.m.\ correction are included in the calculation of the total energy only but not in the variational equations.
\item : No c.m.\ correction.
\item : The one-body c.m.\ correction only is taken into account in the variational equations and is included in the energies
       printed at the end of the iterations.
\item : Two-body c.m.\ correction included self consistently.
\end{enumerate}

{\sf nfunc}:
\begin{enumerate}
\addtocounter{enumi}{-1}
\item Coupling constants of the EDF
calculated from the $t$, $x$, and $W$ of the central, tensor,
and spin-orbit Skyrme force.
\item Coupling constants of the EDF are not related to a Skyrme force.
\end{enumerate}

Note that the code determines from the values of the data which tensor terms
have to be calculated. If $t_e$, $t_o$ or $b_{16}$, $b_{17}$ are
set to zero, only the terms coming from the central part of the interaction
are calculated. When $b_{14}$, $b_{15}$ are set to zero in addition to
$b_{16}$ and $b_{17}$ (as is the case for many hybrid Skyrme
functionals), the calculation of the tensor term is completely omitted.
This significantly reduces the computational time.

{\sf nmass}: Treatment of nucleon masses. All numbers provided by the code are consistent with the latest recommendations from the NIST~\cite{Mohr10}.
\begin{enumerate}
\addtocounter{enumi}{-1}
\item Both masses are equal to the average of neutron and proton
\begin{equation*}
(m_n+m_p)/2 = 938.9187125 \, \text{MeV} \, c^{-2}\, .
\end{equation*}
\item Proton and neutron mass are different
\begin{eqnarray*}
m_p & = & 938.272 046 \, \text{MeV} \, c^{-2} \, ,\\
m_n & = & 939.565 379 \, \text{MeV} \, c^{-2} \, .
\end{eqnarray*}
\item Read $\hbar^2/(2m_n)$ and $\hbar^2/(2m_p)$ from data and set $\hbar$
to the experimental value
\begin{equation*}
\hbar = 6.58211928 \times 10^{-22} \text{MeV s}\, .
\end{equation*}
The nucleon masses are then determined accordingly.
\end{enumerate}

{\sf ncoex} : Treatment of Coulomb exchange term in Eq.~\eqref{eq:EF:schematic}.
\begin{enumerate}
\addtocounter{enumi}{-1}
\item Slater approximation is used, as in Eq.~\eqref{eq:Slater}.
\item The Coulomb exchange term is set to zero.
\end{enumerate}
\end{itemize}

\subsection{Interaction in the pairing channel}
\begin{table}[h!]
\centering
\begin{tabular}{|l|l|}
\hline
   \multicolumn{2}{|c|}{\textbf{Pairing Interaction}}\\
  \hline
  {\sf npair,icut}                    &  2i5    \\
  {\sf gn,encut,delmax(1),alpha}      &  4e15.8 \\
  {\sf gp,epcut,delmax(2)      }      &  3e15.8 \\
  \hline
\end{tabular}
\end{table}

\begin{itemize}
\item {\sf npair} defines the method used to determine the occupation probabilities of
      the single-particle orbitals
      \begin{center}
      \begin{tabular}{l}
           0 - Hartree-Fock,          \\
           1 - BCS       with seniority pairing force,   \\
           2 - BCS       with fixed pairing gaps,         \\
           3 - BCS + LN  with seniority pairing force,    \\
           4 - BCS       with delta pairing force,        \\
           5 - BCS + LN  with delta pairing force.        \\
      \end{tabular}
      \end{center}
      These options are explained in more detail in Sect.~\ref{sect:pairing}.

\item {\sf icut}: Determines the cutoff procedure for the pairing equations.
\begin{enumerate}
 \addtocounter{enumi}{-1}
 \item Cutoff above the Fermi energy. The cutoff function is taken as defined in Eq.~\eqref{eq:pair:cutoff:1}.
 \item Cutoff above and below the Fermi energy. The cutoff function is taken as defined in Eq.~\eqref{eq:pair:cutoff:2}.
\end{enumerate}
\item {\sf gn, gp:} Strengths $G_n, G_p$ of the pairing matrix element in the case of seniority pairing (\textsf{npair=1,3}). Intensities $V_n$, $V_p$ of the pairing interactions in units of MeV fm$^3$, between neutrons and protons respectively in the case of delta pairing (\textsf{npair=4,5}). See Sect. \ref{sect:pairing}.
\item {\sf alpha:} Governs the density dependence of the pairing
strength in Eq.~(\ref{eq:vpair:ULB}).\\
\begin{equation}
\left\{
\begin{array}{l}
\displaystyle V_n = g_n \, \big[ 1 - \alpha \, \rho(r)/\rho_s \big]\, , \\[1mm]
\displaystyle V_p = g_p \, \big[1 - \alpha \, \rho(r)/\rho_s \big]\, .
\end{array}
\right.
\end{equation}

\item {\sf encut,epcut}: Determine the distance from the Fermi energy at which the pairing
interactions are cut, that is, the parameters $\Delta \epsilon_{n}$ and $\Delta \epsilon_{p}$
in Eqs.~(\ref{eq:pair:cutoff:1}) and (\ref{eq:pair:cutoff:2}).

\item {\sf delmax(1), delmax(2):} When using a pairing force with constant pairing gaps ({\sf npair}
equal to 2), these are the values of the neutron ($\Delta_n$) and proton ($\Delta_p$) gaps
respectively. When using other pairing interactions, these data are read, but not used by the code.

\end{itemize}

\subsection{Convergence Thresholds}
  \begin{table}[h!]
  \centering
  \begin{tabular}{|l|l|}
  \hline
  \multicolumn{2}{|c|}{\textbf{Convergence Thresholds}}\\
  \hline
  {\sf epse,epsdh,epsq,epsf}      &   4e15.8 \\
  \hline
  \end{tabular}
  \end{table}

  \begin{itemize}
   \item{epse:} Tolerated relative change of the total energy between mean-field iterations.
   \item{epsdh:} Tolerated value of the weighted sum of the dispersions of the single-particle states.
   \item{epsq:} Tolerated change of the constrained multipole moments between mean-field iterations.
   \item{epsf:} Tolerated change of the Fermi energies of protons and neutrons between mean-field iterations.
  \end{itemize}

\subsection{Density shape constraining parameters}\label{constraintdata}
 \begin{table}[h!]
 \centering
 \begin{tabular}{|l|l|}
 \hline
  \multicolumn{2}{|c|}{\textbf{ Shape Constraints}}\\
  \hline
  {\sf imtd,ifrt,imtg,icutq}          &  4i5    \\
  {\sf ral,epscst,rcut }              &  3e15.8 \\
  {\sf cqr,rtcst or cqr,rncst,rpcst}  &  3e15.8 \\
  {\sf cq2,delq}                      &  2e15.8 \\
  {\sf q1t,q2t or q1n,q2n,q1p,q2p }   &  4e15.8 \\
  \hline
  \end{tabular}
\end{table}
\begin{itemize}

\item {\sf imtd:} defines whether the constraints are put on the total density or on the nucleon densities separately and whether or not the readjustment described in Eq.~\eqref{eq:ReadjustmentConstraints} is applied.
\begin{enumerate}
 \addtocounter{enumi}{-1}
 \item constraint on the total density, no readjustment.
 \item constraint on the total density, readjustment.
 \item constraint on neutron and proton densities independently, no readjustment.
 \item constraint on neutron and proton densities independently, readjustment.
\end{enumerate}

\item {\sf ifrt:} defines the value of the constraint at the first iteration.
      If set to 0, the initial constraint $\mu$ is read from the data; if set to
      1, the value of $\mu$ is read from the file {\sf fort.12}.
      This option should be used for the continuation of a previous run.

\item {\sf imtg:}
\begin{enumerate}
\addtocounter{enumi}{-1}
\item Quadrupole constraint on $q$ and $\gamma$.
\item Quadrupole constraint on $q$ only, $\gamma$ varies freely.
\end{enumerate}

\item {\sf icutq:} Determines the type of cutoff procedure generated
for the constraints.
\begin{enumerate}
\addtocounter{enumi}{-1}
\item  Density-dependent cutoff as described by Eq.~\eqref{eq:cutoff:0} and in \cite{rutz95a}.
\item  Density-dependent cutoff as described by Eq.~\eqref{eq:cutoff:1}.
\item  Spherical, density-independent cutoff as described by Eq.~\eqref{q2}.
\end{enumerate}
See Sect.~\ref{sect:CutoffMultipole} for details.
\item {\sf ral:} Determines the damping of the mean-field potential generated by the constraints. When it is equal zero, it is reset to 0.1.

\item {\sf epscst} is a slow-down factor for the readjustment of the
      constraint, see Eq.~\eqref{eq:ReadjustmentConstraints}. If {\sf epscst} is equal to 0, it is set to 0.02.

\item {\sf rcut:} Determines the radius of the cutoff procedure used for the mean-field potential generated by the constraints. See Sect. \ref{sect:CutoffMultipole} for details.

\item {\sf cqr:} Determines the intensity of the monopole constraint. In absence of such a constraint, it should be set to zero.

\item {\sf rtcst} or {\sf rncst,rpcst} defines the value $\mu$ (or $\mu_n$ and $\mu_p$) for the monopole constraint.

\item {\sf delq} is the unit for the quadrupole deformation mesh
$\Delta q$; it must be
      strictly positive (see {\sf q1t,q2t} below).

\item {\sf cq2} is the intensity of the quadrupole constraint. It corresponds to
      the constant $C$ in Eq.~\eqref{constraint}. {\sf cq2} should be positive or zero.

\item {\sf q1t,q2t} or {\sf q1n,q2n,q1p,q2p} define the quadrupole
      constraint in units of {\sf delq}, in the $q_1, q_2$ representation of Sect. \ref{QuadruSection}. {\sf q1} and {\sf q2} can be
      positive, zero or negative.

\end{itemize}

\section{Output}

The output is subdivided into three parts: a summary of the input,
a limited printout of the results throughout the iterations, and a detailed printout at the end of the
run.
The printing of the input covers the essential information about the wave function from which the calculation is started (read
from {\sf fort.12}) and the data used to run the code. A limited printout of the properties of the input wave functions at zero iterations is also
provided at this stage. Obviously, the value of $d2(h)$ is not yet meaningful at this point. The remaining part of the output is self-explanatory as
it closely follows the notation of the quantities introduced and explained in the article.

\section{Auxiliary codes}

We provide the same auxiliary codes (\texttt{Nil8, Int8, Den8}) as in our first paper as complementary material. They have been updated to read the wave functions generated by the new version of \texttt{Ev8}. These codes still operate in the same way as explained in \cite{BFH05a}.

\section*{Acknowledgements}

The first collaborators with whom this code has been developed have unfortunately left the field. We want to mark our deep gratitude in particular to the late Paul Bonche and to Hubert Flocard, whose expertise in mean-field methods and their irreplaceable competence have allowed us to set up this code. Many other collaborators have also contributed to this work by their remarks, their constructive criticism and sometimes their involvement in coding, in particular, J. Meyer, J. Dobaczewski, B. Gall, N. Tajima, J. Terasaki, B. Avez and B. Bally. This work has been supported in part by the European Union's Seventh Framework Programme ENSAR under grant agreement 262010, by the Belgian Office for Scientific Policy under Grant No. PAI-P7-12 and by the CNRS/IN2P3 through PICS No.\ 5994.
V.H. acknowledges financial support from the F.R.S.-FNRS Belgium.
Part of the computations were performed using the Plateforme Technologique de Calcul Intensif (PTCI) located at the University of Namur, Belgium, which is supported by the F.R.S.-FNRS under the convention No. 2.4520.11. The PTCI is member of the Consortium des \'Equipements de Calcul Intensif (C\'ECI).
Another part of the computations were performed on the HPC cluster HYDRA hosted at the Computing Centre, an IT service co-funded by the Vrije Universiteit Brussel and the Universit\'e Libre de Bruxelles.

\clearpage

\appendix
\section{List of predefined parametrizations}
\label{sec:PreDef}
Here, we list the parametrizations that are predefined in the \texttt{Ev8} code.

\begin{table}[h!]
\centering
\begin{tabular}{|l|l|l|}
  \hline
  Parametrization & \texttt{Ev8} Keyword & Reference \\
  \hline
  Skm* & Skm* & \cite{Bar82a} \\
  Skm  & Skm &  \cite{Kri80a}  \\
  SIII & SIII & \cite{Bei75a} \\
  Ska & Ska &  \cite{Kohl76}\\
  SGII & SGII & \cite{VanG81}\\
  SLyIII$_{0.7}$ & SLYIII0.7 & \cite{washiyama12a} \\
  SLyIII$_{0.8}$ & SLYIII0.8 & \cite{washiyama12a}\\
  SLyIII$_{0.9}$ & SLYIII0.9 & \cite{washiyama12a}\\
  SLyIII$_{1.0}$ & SLYIII1.0 & \cite{washiyama12a}\\
  SLy4 & SLy4 & \cite{Cha98a}\\
  SLy5 & SLy5 & \cite{Cha98a}\\
  SLy6 & SLy6 & \cite{Cha98a}\\
  SLy7 & SLy7 & \cite{Cha98a}\\
  SkP  & SkP &  \cite{dob84}\\
  SkI3 & SkI3 &\cite{reinhard95a}\\
  SkI4 & SkI4 & \cite{reinhard95a}\\
  T22 & T22 & \cite{les07a} \\
  T24 & T24 & \cite{les07a} \\
  T26 & T26&  \cite{les07a} \\
  T42 & T42 &\cite{les07a} \\
  T44 & T44& \cite{les07a} \\
  T46 & T46& \cite{les07a} \\
  T62 & T62& \cite{les07a} \\
  T64 & T64& \cite{les07a} \\
  T66 & T66& \cite{les07a} \\
  SkX & SkX& \cite{bro98}\\
  SLy4T & SLy4T  & \cite{zal08} \\
  SLy5+T & SLy5T & \cite{colo07}\\
  SV-Min & SVMin & \cite{klu09a}\\
  UNEDF0 & UNEDF0& \cite{unedf0} \\
  UNEDF1 & UNEDF1& \cite{unedf1}\\
  UNEDF1SO & UNEDF1SO& \cite{shi14}\\
  \hline
 \end{tabular}
\end{table}

Note that most of these parametrizations have been constructed for use in the mean-field channel only, a separate functional having to be adjusted for the pairing channel.
However, for some parametrizations such as the UNEDF family, both the mean-field and the pairing channel have been adjusted simultaneously. Then, the pairing
part of that EDF needs to be treated with care, because the value of the cut-off above the Fermi energy is not equivalent in different numerical schemes. A cutoff determined for an harmonic oscillator basis will not result in the same pairing correlations in \texttt{Ev8}. Firstly, the properties of the single-particle wave functions above the Fermi energy on a mesh or in a harmonic oscillator basis are completely different. Therefore a cutoff energy has to be defined in concordance with the basis. A second issue is that pairing correlations in \texttt{Ev8} are treated at the BCS level of approximation, which is not valid when single-particle states are in the continuum.

\section{Walltime}
\label{sec:Walltime}

\begin{table}
\begin{center}
 \begin{tabular}{|l|l|l|l|}
  \hline
  nx & SLy4 & T26 & SLy6 \\
  \hline
  20 & 506 & 612 & 1534 \\
  24 & 1043& 1596& 2975 \\
  28 & 1902& 2065& 5139 \\
  32 & 3863& 4412& 9176\\
  \hline
 \end{tabular}
 \end{center}
\caption{\label{tab:Timing}
Time in seconds taken by the processors on the Hercules cluster for 1000 iterations without pairing on the nucleus $^{208}$Pb for different parametrizations and number of mesh points.}
\end{table}

We present in Table \ref{tab:Timing} the time needed to perform 1000 iterations for the nucleus $^{208}$Pb without pairing, for different parametrizations and box parameters. The compiler used was \textsc{pgf90}, version 11.4, with options '-O3 -fastsse' on the \textsc{Hercules}
cluster at UNAMUR. These timings are indicative but probably not very accurate,  for two reasons: firstly, some optimisations performed by the compiler are likely to depend on the size of the mesh and secondly, we have no control on how the computer cluster allocates resources to different jobs.

While in general the code performance is highly dependent on a large number of parameters such as the compiler version and options, some trends are clearly visible. First of all, the time required per iteration grows very quickly when more points are present in the box. Secondly, more `complicated' parametrizations take more time. The T26 parametrization takes in general slightly more walltime than the SLy4 parametrization, due to the presence of the tensor terms. The SLy6 parametrization takes significantly more time than the other two, due to the presence of the two-body centre-of-mass correction.
The additional computational time needed when increasing the number of mesh points or changing to a parametrization with more expensive options, however, does not follow simple scaling laws as the level of optimization achieved by the compiler sensitively depends on the array size of the coordinate space functions.

In Table \ref{tab:Breakdown} we present an indicative breakdown of computing times for a typical calculation with the SLy4 parametrization and zero-range pairing. The most costly operation is obviously the computation of the derivatives (including the construction of the action of the single-particle Hamiltonian) on the mesh. This computation accounts for approximately two thirds of computation time, and roughly speaking all other computation times are negligible in comparison.

\begin{table}
\centering
 \begin{tabular}{|l|l|}
  \hline
  Calculation of \ldots & Total time \\
  \hline
  Derivatives & 66 \%\\
  Densities   & 13 \% \\
  Pairing matrix elements & 10 \%\\
  Orthogonalisation & 7 \%\\
  Coulomb potential & 2 \%\\
  \hline
 \end{tabular}
 \caption{Indicative breakdown of computational cost for a typical calculation with the SLy4 parametrization and zero-range pairing.}
 \label{tab:Breakdown}
\end{table}

\clearpage


\begin{thebibliography}{100}

	\bibitem{bender03a} M. Bender, P.-H. Heenen and P.-G. Reinhard,
Rev. Mod. Phys. \textbf{75}, 121 (2003).

	\bibitem{blaizotripka} J. P. Blaizot and G. Ripka, \textit{Quantum Theory of Finite Systems}
(MIT Press, Cambridge, MA, 1986).

	\bibitem{RingSchuck} P. Ring and P. Schuck,
\textit{The Nuclear Many-Body Problem}
(Springer-Verlag, New York, USA, 1980).

	\bibitem{erler11a} J. Erler, P. Kl{\"u}pfel and P.-G. Reinhard,
J. Phys. G \textbf{38}, 033101 (2011).

	\bibitem{sadoudi13a} J. Sadoudi, T. Duguet, J. Meyer and M. Bender,
Phys. Rev. C \textbf{88}, 064326 (2013).

	\bibitem{lac09a} D. Lacroix, T. Duguet and M. Bender,
Phys. Rev. C \textbf{79}, 044318 (2009).

	\bibitem{BFH05a} P. Bonche, H. Flocard and P.-H. Heenen,
Comp. Phys. Comm. \textbf{171} 49 (2005).

	\bibitem{BFH85a} P. Bonche, H. Flocard, P.-H. Heenen, S. J. Krieger and M. S. Weiss,
Nucl. Phys. \textbf{A443}, 39 (1985).

	\bibitem{BH08a} M. Bender and P.-H. Heenen,
Phys. Rev. C \textbf{78}, 024309 (2008).

	\bibitem{BBH08a} M. Bender, G. F. Bertsch and P.-H. Heenen,
Phys. Rev. C \textbf{78}, 054312 (2008).

	\bibitem{cusson76a} R. Y. Cusson, R. K. Smith and J. A. Maruhn,
Phys. Rev. Lett. \textbf{36}, 1166 (1976).

	\bibitem{flocard78a} H. Flocard, S. E. Koonin and M. S. Weiss,
Phys. Rev. C \textbf{17}, 1682 (1978).

	\bibitem{bonche78a} P. Bonche, B. Grammaticos and S. Koonin,
Phys. Rev. C \textbf{17}, 1700 (1978).

	\bibitem{maruhn} J. A. Maruhn, P.-G. Reinhard, P. D. Stevenson and A. S. Umar,
Comp. Phys. Comm. \textbf{85}, 1410 (2014).

	\bibitem{simenel12a} C. Simenel,
Eur. Phys. J. A \textbf{48}, 152 (2012).

	\bibitem{HFODD1} J. Dobaczewski and J. Dudek,
Comp. Phys. Comm. \textbf{102}, 166 (1997).

	\bibitem{HFODD8} N. Schunck, J. Dobaczewski, J. McDonnell, W. Satu{\l}a, J. A. Sheikh,
A. Staszczak, M. Stoitsov and P. Toivanen,
Comp. Phys. Comm. \textbf{183}, 166 (2012).

	\bibitem{HFBTHO1} M. V. Stoitsov, J. Dobaczewski, W. Nazarewicz and P. Ring,
Comp. Phys. Comm. \textbf{167}, 43 (2005).

	\bibitem{HOSPHE1} B. G. Carlsson, J. Dobaczewski, J. Toivanen and P. Vesel{\'y},
Comp. Phys. Comm. \textbf{181}, 1641 (2010).

	\bibitem{reinhard91a} P.-G. Reinhard,
in \textit{Computational Nuclear Physics 1
-- Nuclear Structure},
K. Langanke, J. A. Maruhn and S. E. Koonin (edts.),
Springer, Berlin, p. 28.

	\bibitem{HFBRAD1} K. Bennaceur and J. Dobaczewski,
Comp. Phys. Comm. \textbf{168}, 96 (2010).

	\bibitem{Karim} K. Bennaceur, \textit{Lenteur HFB Code}, unpublished

	\bibitem{blum92a} V. Blum, G. Lauritsch, J. A. Maruhn and P.-G. Reinhard,
J. Comput. Phys. \textbf{100}, 364 (1992).

	\bibitem{BH86a} D. Baye and P.-H. Heenen,
J. Phys. \textbf{A19}, 2041 (1986).

	\bibitem{DVR} R. G. Littlejohn, M. Cargo, T. Carrington Jr., K. A. Mitchell,
and B. Poirier,
J. Chem. Phys. \textbf{116}, 8691 (2002).

	\bibitem{bulgac13a} A. Bulgac and M. McNeil Forbes,
Phys. Rev. C \textbf{87}, 051301 (2013).

	\bibitem{VHB00a} A. Valor, P.-H. Heenen and P. Bonche,
Nucl. Phys. \textbf{A671}, 145 (2000).

	\bibitem{rutz95a} K. Rutz, J. A. Maruhn, P.-G. Reinhard and W. Greiner,
Nucl. Phys. \textbf{A590}, 680 (1995).


	\bibitem{dob00a} J. Dobaczewski, J. Dudek, S. G. Rohozi{\'n}ski and T. R. Werner,
Phys. Rev. C \textbf{62}, 014310 (2000).

	\bibitem{hellemans12a} V. Hellemans, P.-H. Heenen and M. Bender,
Phys. Rev. C \textbf{85}, 014326 (2012).

	\bibitem{dob00b} J. Dobaczewski, J. Dudek, S. G. Rohozi{\'n}ski and T. R. Werner,
Phys. Rev. C \textbf{62}, 014311 (2000).

	\bibitem{per04a} E. Perli{\'n}ska, S. G. Rohozi{\'n}ski, J. Dobaczewski and W. Nazarewicz,
Phys. Rev. C \textbf{69}, 014316 (2004).

	\bibitem{ben09a} M. Bender, K. Bennaceur, T. Duguet, P.-H. Heenen, T. Lesinski and J. Meyer,
Phys. Rev. C \textbf{80}, 064302 (2009).

	\bibitem{Nilsson} S. G. Nilsson and I. Ragnarsson,
\textit{Shapes and Shells in Nuclear Structure}
(Cambridge University Press, Cambridge, England, 1995).

	\bibitem{Rowe70a} D. J. Rowe,
\textit{Nuclear Collective Motion}
(Methuen, London, 1970).

	\bibitem{Bei75a} M. Beiner, H. Flocard, N. Van Giai and P. Quentin,
Nucl. Phys. \textbf{A238}, 29 (1975).

	\bibitem{Kri80a} H. Krivine, J. Treiner and O. Bohigas,
Nucl. Phys. {\bf A336}, 155 (1980).

	\bibitem{Bar82a} J. Bartel, P. Quentin, M. Brack, C. Guet and H.-B. H{\aa}kansson,
Nucl. Phys. \textbf{A386}, 79 (1982).

	\bibitem{vautherin72a} D. Vautherin and D. M. Brink,
Phys. Rev. C \textbf{5}, 626 (1972).

	\bibitem{bell56a} J. S. Bell and T. H. R. Skyrme,
Phil. Mag. \textbf{1}, 1055 (1956);
T. H. R. Skyrme,
Nucl. Phys. \textbf{9}, 635 (1958).

	\bibitem{sky56a} T. H. R. Skyrme,
Phil. Mag. \textbf{1}, 1043 (1956);
Nucl. Phys. \textbf{9}, 615 (1958).

	\bibitem{cochet04a} B. Cochet, K. Bennaceur, P. Bonche, T. Duguet and J. Meyer,
Nucl. Phys. \textbf{A731}, 34 (2004).

	\bibitem{lesinski06a} T. Lesinski, K. Bennaceur, T. Duguet and J. Meyer,
Phys. Rev. C \textbf{74}, 044315 (2006).

	\bibitem{les07a} T. Lesinski, M. Bender, K. Bennaceur, T. Duguet and J. Meyer,
Phys. Rev. C \textbf{76}, 014312 (2007).

	\bibitem{bonche87} P. Bonche, H. Flocard and P.-H. Heenen,
Nucl. Phys. \textbf{A467}, 115 (1987).

	\bibitem{dob96d} J. Dobaczewski and J. Dudek,
Acta Phys. Pol. B \textbf{27}, 45 (1996).

	\bibitem{dob84} J. Dobaczewski, H. Flocard and J. Treiner, Nucl. Phys. \textbf{A422}, 103 (1984).

	\bibitem{gomex92a} J. M. G. G{\'o}mex, C. Prieto and J. Navarro,
Nucl. Phys. \textbf{A549}, 125 (1992).

	\bibitem{pot10a} K. J. Pototzky, J. Erler, P.-G. Reinhard and V. O. Nesterenko,
Eur. Phys. J. A \textbf{46}, 299 (2010).

	\bibitem{schunck10b} N. Schunck J. Dobaczewski, J. McDonnell, J. Mor{\'e}, W. Nazarewicz,
J. Sarich and M.~V. Stoitsov, Phys. Rev. C {\bf 81}, 024316 (2010).

	\bibitem{hellemans13a} V. Hellemans, A. Pastore, T. Duguet, K. Bennaceur, D. Davesne,
J. Meyer, M. Bender and P.-H. Heenen,
Phys. Rev. C \textbf{88}, 064323 (2013).

        \bibitem{cao10a} L.-G. Cao, G. Col{\`o} and H. Sagawa, Phys. Rev. C \textbf{81}, 044302 (2010).

	\bibitem{Cha98a} E. Chabanat, P. Bonche, P. Haensel, J. Meyer and R. Schaeffer,
Nucl. Phys. \textbf{A635}, 231 (1998);
Nucl. Phys. \textbf{A643}, 441(E) (1998).

	\bibitem{chamel08a} N. Chamel, S. Goriely and J. Pearson,
Nucl. Phys. \textbf{A812}, 72 (2008).

	\bibitem{unedf1} M. Kortelainen, J. McDonnell, W. Nazarewicz, P.-G. Reinhard,
J. Sarich, N. Schunck, M. V. Stoitsov and S. M. Wild,
Phys. Rev. C \textbf{85}, 024304 (2012).

	\bibitem{unedf2} M. Kortelainen, J. McDonnell, W. Nazarewicz, E. Olsen, P.-G. Reinhard,
J. Sarich, N. Schunck, S. M. Wild, D. Davesne, J. Erler and A. Pastore,
Phys. Rev. C \textbf{89}, 054314 (2014).

	\bibitem{Mohr10} P. J. Mohr, B. N. Taylor and D. B. Newe,
Rev. Mod. Phys. \textbf{84}, 1527 (2012).

	\bibitem{anguiano01a} M. Anguiano, J. L. Egido and L. M. Robledo,
Nucl. Phys. \textbf{A683}, 227 (2001).

	\bibitem{lebloas11a} J. Le Bloas, M.-H. Koh, P. Quentin, L. Bonneau and J. A. Ithnin,
Phys. Rev. C \textbf{84}, 014310 (2011).

	\bibitem{ben00b} M. Bender, K. Rutz, P.-G. Reinhard and J. A. Maruhn,
Eur. Phys. J. \textbf{A7}, 47 (2000).

	\bibitem{reinhard95a} P.-G. Reinhard and H. Flocard,
Nucl. Phys. \textbf{A584}, 467 (1995).

	\bibitem{klu09a} P. Kl{\"u}pfel, P.-G. Reinhard, T. J. B{\"u}rvenich and J. A. Maruhn,
Phys. Rev. C \textbf{79}, 034310 (2009).

	\bibitem{kim97a} K.-H. Kim, T. Otsuka and P. Bonche,
J. Phys. G \textbf{23}, 1267 (1997).

	\bibitem{bro98} B. A. Brown, Phys. Rev. C \textbf{58}, 220 (1998).

	\bibitem{dob96a} J. Dobaczewski, W. Nazarewicz, T. R. Werner, J. F. Berger, C. R. Chinn,
and J. Decharg{\'e},
Phys. Rev. C \textbf{53}, 2809 (1996).

	\bibitem{blo76a} J. B{\l}ocki and H. Flocard,
Nucl. Phys. \textbf{A273}, 45 (1976).

	\bibitem{MBD91a} J. Meyer, P. Bonche, J. Dobaczewski, H. Flocard and P.-H. Heenen, Nucl. Phys. {\bf A533} (1991) 307.

	\bibitem{HVB01a} P.-H. Heenen, A. Valor, M. Bender, P. Bonche and H. Flocard, Eur. Phys. J. {\bf A11} (2001) 393.

	\bibitem{krieger90a} S. J. Krieger, P. Bonche, H. Flocard, P. Quentin and M. S. Weiss,
Nucl. Phys. \textbf{A517}, 275 (1990).

	\bibitem{ter95a} J. Terasaki, P-H Heenen, P. Bonche, J. Dobaczewski and H. Flocard,
Nucl. Phys. {\bf A593} (1995) 1-20.

	\bibitem{rig99a} C. Rigollet, P. Bonche, H. Flocard and P.-H. Heenen,
Phys. Rev. C \textbf{59}, 3120 (1999).

	\bibitem{ben00a} M. Bender, K. Rutz, P.-G. Reinhard and J. A. Maruhn,
Eur. Phys. J. \textbf{A8}, 59 (2000).

	\bibitem{cwiok96} S. {\'C}wiok, J. Dobaczewski, P.-H. Heenen, P. Magierski and W. Nazarewicz,
Nucl. Phys. \textbf{A611}, 211 (1996).

	\bibitem{dob01a} J. Dobaczewski, W. Nazarewicz and P.-G. Reinhard,
Nucl. Phys. \textbf{A693}, 361 (2001).

	\bibitem{bertsch09a} G. F. Bertsch, C. A. Bertulani, W. Nazarewicz, N. Schunck and M. V. Stoitsov,
Phys. Rev. C \textbf{79}, 034306 (2009).

	\bibitem{yu03a} A. Bulgac and Y. Yu, Phys. Rev. Lett. \textbf{88}, 042504 (2002).

	\bibitem{quentin90a} P. Quentin, N. Redon, J. Meyer and M. Meyer,
Phys. Rev. C \textbf{41}, 341 (1990);
Phys. Rev. C \textbf{43}, 361(E) (1991).

	\bibitem{bennour89a} L. Bennour, P.-H. Heenen, P. Bonche, J. Dobaczewski and H. Flocard, Phys. Rev. C \textbf{40}, 2834 (1989).

	\bibitem{rei96a} P.-G. Reinhard, W. Nazarewicz, M. Bender and J. A. Maruhn, Phys. Rev. C \textbf{53}, 2776 (1996).

        \bibitem{Myers69} W. D. Myers and W. J. Swiatecki, Ann. Phys. (NY) \textbf{55}, 395 (1969).

	\bibitem{raman01a} S. Raman, C. W. Nestor and Jr. P. Tikkanen,
Atom. Data Nucl. Data Tables \textbf{78}, 1 (2001).

	\bibitem{Hasse88} R. W. Hasse and W. D. Myers,
\textit{Geometrical Relationships of Macroscopic Nuclear Physics},
(Springer-Verlag, Berlin, 1988).

	\bibitem{Sta10} A. Stasczak, M. Stoitsov, A. Baran and W. Nazarewicz, Eur. Phys. J. A \textbf{46}, 85 (2010).

\bibitem{Gal94} B. Gall, P. Bonche, J. Dobaczewski, H. Flocard and P-H Heenen, Z. Phys. A \textbf{348}, 183 (1994).

	\bibitem{Flo73a} H. Flocard, P. Quentin, A. K. Kerman and D. Vautherin,
Nucl. Phys. \textbf{A203}, 433 (1973).

	\bibitem{Bass75} W. H. Bassichis, M. R. Strayer and M. T. Vaughn, Nucl. Phys. \textbf{A263}, 379 (1975).

	\bibitem{Font75} G. Fonte and G. Schiffrer, Nucl. Phys. \textbf{A259}, 20 (1975).

	\bibitem{RBH14} W. Ryssens, M. Bender and P.-H. Heenen, \textit{bo be published}.

	\bibitem{Dav80} K. T. R. Davies, H. Flocard, S. Krieger and M. S. Weiss,
Nucl. Phys. \textbf{A342}, 111 (1980).

	\bibitem{Kohl76} H. S. K\"ohler, Nucl. Phys. \textbf{A258}, 301 (1976).

	\bibitem{VanG81} N. Van Giai and H. Sagawa, Phys. Lett. B \textbf{106}, 5 (1981).

	\bibitem{washiyama12a} K. Washiyama, K. Bennaceur, B. Avez, M. Bender, P.-H. Heenen and V. Hellemans,
Phys. Rev. C \textbf{86}, 054309 (2012).

	\bibitem{zal08} M. Zalewski, J. Dobaczewski, W. Satu{\l}a and T. R. Werner, Phys. Rev. C \textbf{77}, 024316 (2008).

	\bibitem{colo07} G. Col\`o, H. Sagawa, S. Fracasso and P. F. Bortignon, Phys. Lett. B \textbf{646}, 227 (2007);
Phys. Lett. B \textbf{668}, 457(E) (2008).

	\bibitem{unedf0} M. Kortelainen, T. Lesinski, J. More, W. Nazarewicz, J. Sarich,
N. Schunck, M. V. Stoitsov and S. Wild,
Phys. Rev. C \textbf{82}, 024313 (2010).

	\bibitem{shi14} Y. Shi, J. Dobaczewski and P. T. Greenlees, Phys. Rev. C \textbf{89}, 034309 (2014).

	\bibitem{Stoitsov12} M. V. Stoitsov, N. Schunck, M. Kortelainen, N. Michel, H. Nam, E. Olsen, J. Sarich and S. Wild, Comp. Phys. Comm. \textbf{184}, 1592 (2014)
	
	
	
\end{thebibliography}
\end{document}